\documentclass[letter, journal]{IEEEtran}

\ifCLASSOPTIONcompsoc
  \usepackage[nocompress]{cite}
\else
  \usepackage{cite}
\fi
\ifCLASSINFOpdf
\else
\fi

\usepackage{booktabs}
\usepackage{multirow}
\usepackage{forest}
\usepackage{tikz}
\usetikzlibrary{shapes, arrows, positioning}
\usepackage{adjustbox}
\usepackage{hyperref}
\usepackage{pdflscape}
\usepackage{afterpage}
\usepackage{capt-of}
\usepackage{enumitem}
\DeclareUnicodeCharacter{0301}{\'{a}}

\begin{document}
%
\title{A Survey of Multimodal Information Fusion for Smart Healthcare: Mapping the Journey from Data to Wisdom}
%
%
%
%

\author{Thanveer Shaik,  Xiaohui Tao,  Lin Li, Haoran Xie, Juan D. Vel ́asquez
\thanks{Thanveer Shaik, Xiaohui Tao are with 
the School of Mathematics, Physics \& Computing, University of Southern Queensland, Toowoomba, Queensland, Australia (e-mail: Thanveer.Shaik@usq.edu.au, Xiaohui.Tao@usq.edu.au).}
\thanks{Lin Li is with the School of Computer and Artificial Intelligence, Wuhan University of Technology, China (e-mail: cathylilin@whut.edu.cn)}
\thanks{Haoran Xie is with the Department of Computing and Decision Sciences, Lingnan University, Tuen Mun, Hong Kong (e-mail: hrxie@ln.edu.hk)}

\thanks{Juan D. Vel ́asquez is with Industrial Engineering Department, University of Chile, Chile (e-mail: Rajendra.Acharya@usq.edu.au).}
}

\IEEEtitleabstractindextext{%
\begin{abstract}
Multimodal medical data fusion has emerged as a transformative approach in smart healthcare, enabling a comprehensive understanding of patient health and personalized treatment plans. In this paper, a journey from data to information to knowledge to wisdom (DIKW) is explored through multimodal fusion for smart healthcare. We present a comprehensive review of multimodal medical data fusion focused on the integration of various data modalities. The review explores different approaches such as feature selection, rule-based systems, machine learning, deep learning, and natural language processing, for fusing and analyzing multimodal data. This paper also highlights the challenges associated with multimodal fusion in healthcare. By synthesizing the reviewed frameworks and theories, it proposes a generic framework for multimodal medical data fusion that aligns with the DIKW model. Moreover, it discusses future directions related to the four pillars of healthcare: Predictive, Preventive, Personalized, and Participatory approaches. The components of the comprehensive survey presented in this paper form the foundation for more successful implementation of multimodal fusion in smart healthcare. Our findings can guide researchers and practitioners in leveraging the power of multimodal fusion with state-of-the-art approaches to revolutionize healthcare and improve patient outcomes.
\end{abstract}

}

\maketitle

\IEEEdisplaynontitleabstractindextext

%
\IEEEpeerreviewmaketitle

\ifCLASSOPTIONcompsoc
\IEEEraisesectionheading{\section{Introduction}\label{sec:introduction}}
\else
\section{Introduction}
\label{sec:introduction}
\fi

\begin{figure}
    \centering
    \includegraphics[width=\columnwidth]{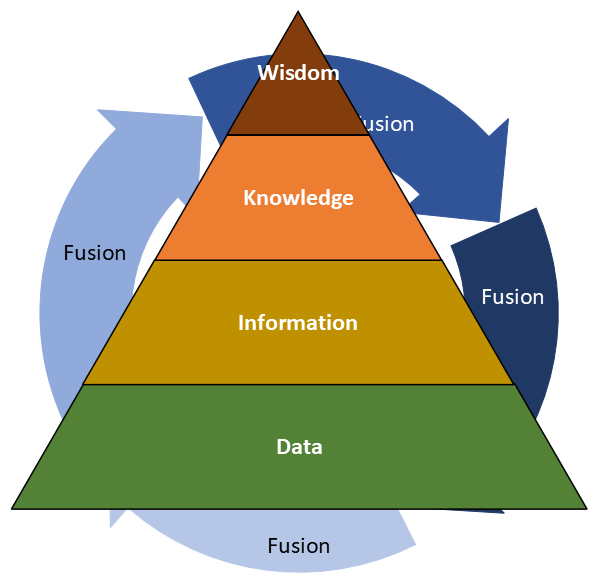}
    \caption{DIKW Fusion conceptual model}
    \label{fig:transform}
\end{figure}

In the realm of smart healthcare, where cutting-edge technologies and data-driven approaches are revolutionizing the field, the integration of multimodal data has emerged as a transformative tool to enhance decision-making and improve patient outcomes. This paper presents a comprehensive exploration of multimodal medical data fusion for smart healthcare, illustrating the journey from raw data to actionable insights through the four-level pyramid shown in Fig.~\ref{fig:transform}.

The Data Information Knowledge Wisdom (DIKW) model is a conceptual framework that illustrates the hierarchical progression of data into wisdom~\cite{ackoff1989data}. Through its process, raw data is transformed into meaningful information, knowledge, and ultimately wisdom, which can be used for informed decision-making and problem-solving~\cite{fiore2018data}. The DIKW model recognizes that data alone is not sufficient to drive insights and actions. Instead, data needs to be processed, organized, and contextualized to extract valuable information. This information is then synthesized and combined with existing knowledge to gain understanding, leading to the development of knowledge. Knowledge, in turn, can then be applied in practical situations to make informed decisions and solve complex problems, resulting in wisdom.

At the base of the pyramid in Fig.~\ref{fig:transform}, we have the Data level, which encompasses diverse sources of data such as Electronic Health Records (EHRs), medical imaging, wearable devices, genomic data, sensor data, environmental data, and behavioral data. This raw data serves as the foundation for subsequent analysis and interpretation. Moving up the pyramid, we reach the Information level, where the raw data undergoes processing, organization, and structuring to derive meaningful and contextualized information. For instance, heart rate data from a wearable device can be processed to determine average resting heart rates, activity levels, and potential anomalies.

The Knowledge level, situated above Information, represents the interconnected structure of the organized data from various sources. By establishing relationships and connections between entities like patients, diseases, or medical treatments, the Knowledge level enables the identification of patterns, trends, and correlations. It facilitates a holistic understanding of the data and serves as a powerful tool for generating insights.

Finally, at the pinnacle of the pyramid, we have the Wisdom level. This is where actionable insights are derived from the Knowledge level, allowing informed decision-making, prediction of future outcomes~\cite{tao2020mining}, and a deeper understanding of complex phenomena. These insights enable personalized treatment plans, predictions about disease progression, and the identification of risk factors.

It's important to point out a key feature of the model, its circular structure, shown by the arrows in Fig.~\ref{fig:transform}. These arrows indicate that combining different types of data assists in the progression of Data to Information, then to Knowledge and Wisdom. This cyclical nature adds flexibility to the model, allowing for constant updates and improvements in how data is processed. In this sense, reaching the level of Wisdom helps to fine-tune the steps and methods used at earlier stages, making future data collection, information gathering, and knowledge creation more effective.

This paper further explores different approaches such as feature selection, rule-based systems, machine learning, deep learning, and natural language processing for multimodal fusion. It also addresses challenges related to data quality, privacy, security, processing, analysis, clinical integration, ethics, and interpretation of results. With its emphasis on the transformative potential of multimodal medical data fusion, this paper sets the stage for future research and advancements in the field of smart healthcare, and paves the way for improved patient care outcomes and personalized healthcare solutions.

The following are the key contributions of this paper:
\begin{itemize}
\item The application and adaptation of the existing DIKW conceptual model to describe the journey of data to information to knowledge to wisdom in the context of multimodality fusion for smart healthcare.
\item A taxonomy that organizes state-of-the-art techniques in multimodality fusion with the DIKW conceptual model.
\item A proposed generic DIKW techniques framework for smart healthcare, that not only highlights the current efforts, but also provides a vision for its future evolution.
\item A review of the challenges and recommended solutions associated with multimodal fusion, informed by the existing DIKW conceptual model and the proposed framework, to guide future research directions.
\end{itemize}

The rest of this paper is organized as follows: Section~\ref{modalities} delves into the representation of data in multimodal applications. Section~\ref{approaches} explores various approaches for integrating information from multiple modalities, then outlines the proposed taxonomy. Section~\ref{challenges} examines the challenges and trends associated with multimodal fusion. Section~\ref{generic} presents a generic framework for multimodal fusion that aligns with the DIKW model. In section~\ref{future}, we outline the future directions for multimodal fusion in smart healthcare, with a particular focus on the 4Ps of healthcare. Finally, the paper concludes in Section~\ref{conclusion}.

\section{Modalities in Smart Healthcare}\label{modalities}

\begin{figure}[!h]
    \centering
    \includegraphics[width=\columnwidth]{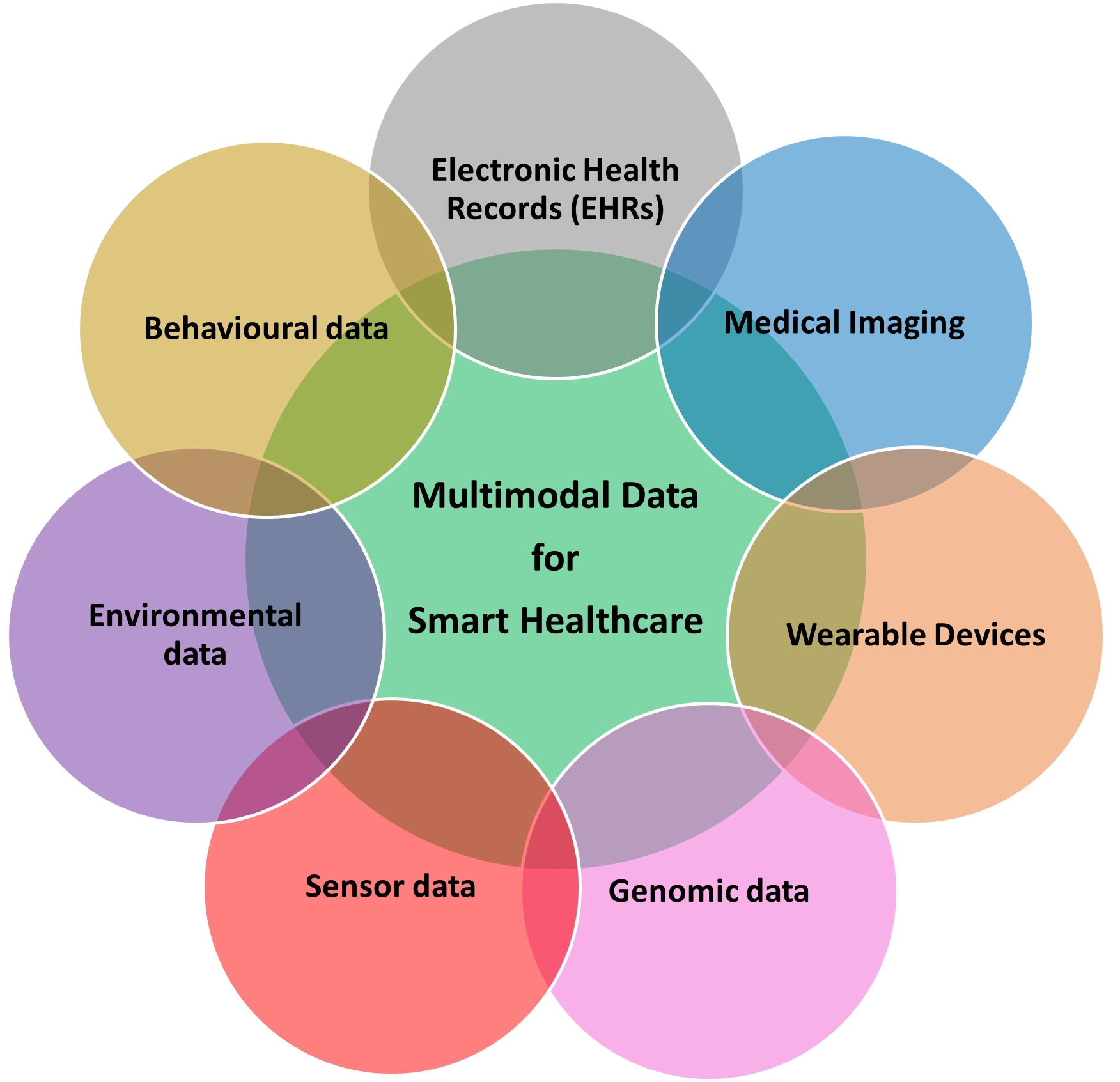}
    \caption{Overview of Multimodality Fusion for Smart Healthcare}
    \label{fig:multimodality}
\end{figure}

There are various data modalities in healthcare, such as EHRs, Medical Imaging, Wearable Devices, Genomic data, Sensor data, Environmental data, and Behavioral data as shown in Fig.~\ref{fig:multimodality}. These modalities contain unstructured raw data specific to their respective formats. As the data is processed, it is transformed into meaningful information through the involvement of techniques such as structuring EHRs, feature extraction from Medical Imaging, and analysis of wearable device data. 

\subsection{Electronic Health Records (EHRs)} 
EHRs serve as a central repository of medical data for healthcare providers, the adoption of which has led to a surge in the amount of intricate patient data~\cite{liang2021adoption}. These datasets, although extensive and tailored to each patient, are often fragmented and may lack organization. They encompass diverse variables like medications, laboratory values, imaging results, physiological measurements, and historical notes~\cite{zhang2020analytics}, which lead to increased complexity in analysis. Machine learning (ML) provides a potential solution to this complexity by enabling the exploration of intricate relationships among the diverse variables present in EHR datasets~\cite{hossain2023use}.

EHRs play a crucial role in multimodal fusion systems within smart healthcare, especially for multidisciplinary and life-threatening diseases like diabetes~\cite{ihnaini2021smart}. However, managing and analyzing unstructured data collected from sensors and EHRs is challenging. Data fusion is crucial for accurate predictions, and deep learning approaches are effective for larger healthcare datasets. Healthcare datasets can be enhanced by collecting patients' data through wearable sensors and EHRs. An ensemble ML approach is employed to develop a recommendation system for accurate prediction and timely recommendations for patients with multidisciplinary diabetes.

To optimize multimodal fusion strategies in EHR data, Xu et al.~\cite{xu2021mufasa} proposed MUFASA, a novel approach that extends Neural Architecture Search (NAS). The authors based their model on the Transformer architecture, which has shown promise in leveraging EHR's internal structure. Experimental results demonstrated that MUFASA outperformed Transformer, Evolved Transformer, RNN variants, and traditional NAS models on public EHR data. MUFASA architectures achieved higher top-5 recall rates compared to Transformer in predicting CCS diagnosis codes, and they outperformed unimodal NAS by customizing modelling for each modality. MUFASA also exhibited effective transfer learning to ICD-9, another EHR task. The representation of EHR data is challenging due to different modalities, such as medical codes and clinical notes, all of which have distinct characteristics. 

Another challenge is the extraction of inter-modal correlations, which are often overlooked or not effectively captured by existing models. An et al.~\cite{an2021main} proposed the Multimodal Attention-based fusIon Networks (MAIN) model, which aims to address two key challenges in healthcare prediction using EHR data. The MAIN model incorporates multiple independent feature extraction modules tailored to each modality, including self-attention and time-aware Transformer for medical codes, and a CNN model for clinical notes. It also introduces an inter-modal correlation extraction module composed of a low-rank multimodal fusion method and a cross-modal attention layer. The model combines the representations of each modality and their correlations to generate visit and patient representations for diagnosis prediction \cite{malakar2022computer}, by leveraging attention mechanisms and neural networks. Overall, MAIN offers a comprehensive framework for multimodal fusion and correlation extraction in EHR-based prediction tasks.

\subsection{Wearable Devices} 
Wearable devices have become increasingly prevalent in the field of smart healthcare, offering the potential to monitor and track various aspects of an individual's health and well-being. These devices, typically worn on the body or incorporated into clothing or accessories, can collect real-time data about vital signs, physical activity, sleep patterns, and other health-related metrics~\cite{papa2020health, teixeira2021wearable}. The data gathered by wearable devices can provide valuable insights into an individual's overall health status, enabling personalized health monitoring and preventive care~\cite{sheth2018will}. Moreover, wearable devices can facilitate remote patient monitoring, allowing healthcare professionals to track patients' health remotely and intervene when necessary. The integration of wearable devices with smart healthcare systems enables continuous monitoring, early detection of health issues, and the ability to deliver personalized interventions and recommendations~\cite{shaik2023remote}.

\subsection{Sensor data} 
Sensor data plays a crucial role in enabling smart healthcare by providing real-time monitoring and tracking of various physiological parameters and activities. Wearable devices, implantable sensors, and remote monitoring systems collect data such as heart rate, blood pressure, temperature, glucose levels, physical activity, sleep patterns, and so on. Sensor data in smart healthcare enables continuous monitoring of an individual's health status, facilitating early detection and intervention for potential health issues~\cite{tao2021remote}. It allows healthcare providers to gather objective and accurate data, leading to more informed decision-making and personalized treatment plans. For example, sensor data can help in managing chronic conditions like diabetes or cardiovascular diseases by monitoring glucose levels or heart rate variability. Real-time sensor data can also enable remote patient monitoring, telemedicine, and telehealth services, allowing healthcare professionals to monitor patients' conditions from a distance~\cite{mohammed2019real}. This is particularly beneficial for individuals with limited mobility or those residing in remote areas, by providing access to healthcare services without the need for frequent hospital visits~\cite{duran2019iot}.

\begin{figure*}
    \centering
    \includegraphics[width=\textwidth]{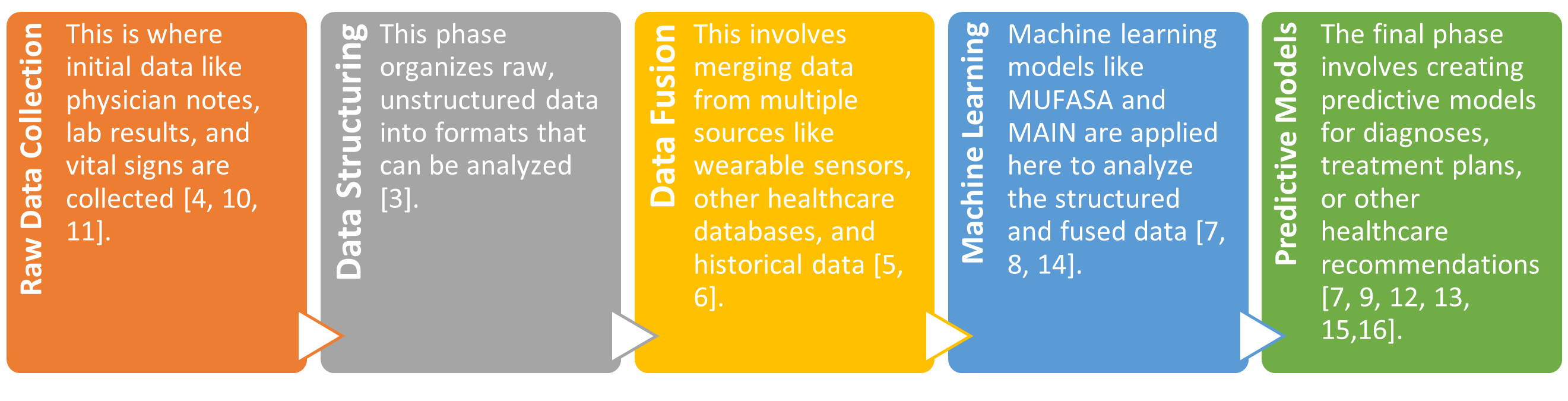}
    \caption{The Lifecycle of Electronic Health Records, Wearable Devices, and Sensors in Smart Healthcare}
    \label{fig:EHR_flowchart}
\end{figure*}

The life cycle for data from EHRs, wearable devices, and sensors in the context of smart healthcare each follow a similar trajectory, encompassing stages such as raw data acquisition, data structuring, data fusion, and ultimately, predictive modeling. This cyclical process is graphically illustrated in Fig.~\ref{fig:EHR_flowchart}.

\subsection{Medical Imaging} 
Medical imaging plays a crucial role in smart healthcare by providing valuable diagnostic information and aiding in the management of various medical conditions~\cite{tian2019smart, senbekov2020recent, linet2012cancer, liu2019comparison, garain2021detection, das2021bi}. It involves the use of advanced imaging technologies to capture detailed images of the human body, allowing healthcare professionals to visualize and analyze anatomical structures, detect abnormalities, and monitor the progress of treatments. In smart healthcare, medical imaging is integrated with digital technologies and data analytics to enhance the efficiency, accuracy, and accessibility of healthcare services~\cite{tian2019smart}.


\subsection{Genomic data}
Genomic data plays a significant role in the realm of smart healthcare, offering valuable insights into an individual's genetic makeup and its impact on their health. This type of data includes information about an individual's DNA sequence, genetic variations, and gene expression patterns~\cite{awotunde2021prediction}. With advancements in genomic sequencing technologies, it has become more accessible and affordable to obtain a person's genetic information. In smart healthcare, genomic data can be utilized for various purposes, such as in the diagnosis and prediction of genetic disorders, as well as in the identification of genetic markers associated with increased disease risk or treatment response~\cite{yu2018five}. Genomic data can also enable personalized medicine by guiding treatment decisions based on an individual's unique genetic profile~\cite{pai2018patient}. For example, it can help determine optimal drug choices and dosages, minimizing adverse reactions and improving treatment outcomes. Integrating genomic data with other health data sources, such as EHRs and wearable devices, can provide a comprehensive overview of an individual's health~\cite{acosta2022multimodal}. This multimodal approach allows for practices such as more accurate assessment of disease risks, personalized prevention strategies, and targeted interventions.

\subsection{Environmental data} 
Environmental data can significantly contribute to smart healthcare by providing insights into the impact of environmental factors on individual health and well-being. Environmental data includes information about air quality, temperature, humidity, pollution levels, noise levels, and other relevant parameters in a person's surroundings. By incorporating environmental data into smart healthcare systems, healthcare providers can better understand the environmental conditions that may influence a person's health outcomes~\cite{taiwo2020smart}. For example, monitoring air quality can help identify areas with high pollution levels, which is particularly valuable for individuals with respiratory conditions like asthma. 

Analyzing this environmental data in real-time allows healthcare professionals to provide personalized recommendations and interventions to mitigate the impact of poor air quality on patients' health~\cite{carlsten2020personal}. Environmental data can also aid in preventive healthcare by identifying patterns and correlations between environmental factors and specific health conditions~\cite{hu2021role, alvarez2020software}. For instance, studying the relationship between temperature and heat-related illnesses facilitates the implementation early warning systems and interventions during heat waves or extreme weather events.

\begin{table*}[!ht]
\centering
\caption{Multimodal Datasets for smart healthcare}
\label{tab:datasets}
\resizebox{\textwidth}{!}{%
\begin{tabular}{@{}lllllll@{}}
\toprule
\textbf{Modality} &
  \textbf{Datatype} &
  \textbf{Dataset} &
  \textbf{No. of Instances} &
  \textbf{No. of Attributes} &
  \textbf{Task} &
  \textbf{Popularity*} \\ \midrule
\multicolumn{1}{l}{\multirow{14}{*}{Single Modality}} &
  \multicolumn{1}{l}{\multirow{2}{*}{EHR}} &
  \multicolumn{1}{l}{\begin{tabular}[c]{@{}l@{}}eICU Collaborative \\ Research Database~\cite{pollard2018eicu}\end{tabular}} &
  \multicolumn{1}{l}{200,000 admissions} &
  \multicolumn{1}{l}{Varies} &
  \multicolumn{1}{l}{\begin{tabular}[c]{@{}l@{}}Various tasks, mainly \\ diagnosis and prognosis\end{tabular}} &
  \multicolumn{1}{l}{Medium} \\ \cmidrule(l){3-7} 
\multicolumn{1}{l}{} &
  \multicolumn{1}{l}{} &
  \multicolumn{1}{l}{MIMIC-III~\cite{johnson2016mimic}} &
  \multicolumn{1}{l}{40,000 patients} &
  \multicolumn{1}{l}{Varies} &
  \multicolumn{1}{l}{\begin{tabular}[c]{@{}l@{}}Various tasks, mainly \\ diagnosis and prognosis\end{tabular}} &
  \multicolumn{1}{l}{High} \\ \cmidrule(l){2-7} 
\multicolumn{1}{l}{} &
  \multicolumn{1}{l}{\multirow{12}{*}{Imaging}} &
  \multicolumn{1}{l}{MRNet~\cite{azcona2020comparative}} &
  \multicolumn{1}{l}{1,370 exams} &
  \multicolumn{1}{l}{MRI data} &
  \multicolumn{1}{l}{Disease detection} &
  \multicolumn{1}{l}{Low} \\ \cmidrule(l){3-7} 
\multicolumn{1}{l}{} &
  \multicolumn{1}{l}{} &
  \multicolumn{1}{l}{\begin{tabular}[c]{@{}l@{}}RSNA Pneumonia \\ Detection Challenge~\cite{shih2019augmenting}\end{tabular}} &
  \multicolumn{1}{l}{30,000 images} &
  \multicolumn{1}{l}{Pneumonia labels} &
  \multicolumn{1}{l}{Disease detection} &
  \multicolumn{1}{l}{Low} \\ \cmidrule(l){3-7} 
\multicolumn{1}{l}{} &
  \multicolumn{1}{l}{} &
  \multicolumn{1}{l}{MURA~\cite{rajpurkar2017mura}} &
  \multicolumn{1}{l}{40,895 images} &
  \multicolumn{1}{l}{Abnormal/normal} &
  \multicolumn{1}{l}{Disease detection} &
  \multicolumn{1}{l}{Medium} \\ \cmidrule(l){3-7} 
\multicolumn{1}{l}{} &
  \multicolumn{1}{l}{} &
  \multicolumn{1}{l}{\begin{tabular}[c]{@{}l@{}}Pediatric Bone Age \\ Challenge Dataset~\cite{halabi2019rsna}\end{tabular}} &
  \multicolumn{1}{l}{Thousands of images} &
  \multicolumn{1}{l}{Bone age} &
  \multicolumn{1}{l}{Bone age estimation} &
  \multicolumn{1}{l}{Medium} \\ \cmidrule(l){3-7} 
\multicolumn{1}{l}{} &
  \multicolumn{1}{l}{} &
  \multicolumn{1}{l}{\begin{tabular}[c]{@{}l@{}}Indiana University Chest \\ X-ray Collection~\cite{demner2016preparing}\end{tabular}} &
  \multicolumn{1}{l}{8,000 images} &
  \multicolumn{1}{l}{\begin{tabular}[c]{@{}l@{}}Chest radiograph \\ DICOM images\end{tabular}} &
  \multicolumn{1}{l}{Various tasks} &
  \multicolumn{1}{l}{Medium} \\ \cmidrule(l){3-7} 
\multicolumn{1}{l}{} &
  \multicolumn{1}{l}{} &
  \multicolumn{1}{l}{FastMRI~\cite{zbontar2018fastmri}} &
  \multicolumn{1}{l}{Thousands of scans} &
  \multicolumn{1}{l}{MRI data} &
  \multicolumn{1}{l}{Image reconstruction} &
  \multicolumn{1}{l}{Medium} \\ \cmidrule(l){3-7} 
\multicolumn{1}{l}{} &
  \multicolumn{1}{l}{} &
  \multicolumn{1}{l}{CheXpert~\cite{irvin2019chexpert}} &
  \multicolumn{1}{l}{224,316 images} &
  \multicolumn{1}{l}{14 labels per image} &
  \multicolumn{1}{l}{Disease detection} &
  \multicolumn{1}{l}{High} \\ \cmidrule(l){3-7} 
\multicolumn{1}{l}{} &
  \multicolumn{1}{l}{} &
  \multicolumn{1}{l}{OASIS Brains Project~\cite{marcus2007open}} &
  \multicolumn{1}{l}{Varies with dataset} &
  \multicolumn{1}{l}{MRI and clinical data} &
  \multicolumn{1}{l}{Brain studies} &
  \multicolumn{1}{l}{High} \\ \cmidrule(l){3-7} 
\multicolumn{1}{l}{} &
  \multicolumn{1}{l}{} &
  \multicolumn{1}{l}{LIDC-IDRI~\cite{armato2011lung}} &
  \multicolumn{1}{l}{Over 1,000 patients} &
  \multicolumn{1}{l}{\begin{tabular}[c]{@{}l@{}}CT scans with marked-up \\ annotated lesions\end{tabular}} &
  \multicolumn{1}{l}{Nodule detection} &
  \multicolumn{1}{l}{High} \\ \cmidrule(l){3-7} 
\multicolumn{1}{l}{} &
  \multicolumn{1}{l}{} &
  \multicolumn{1}{l}{TCIA~\cite{clark2013cancer}} &
  \multicolumn{1}{l}{Millions of images} &
  \multicolumn{1}{l}{Various data types} &
  \multicolumn{1}{l}{Cancer research} &
  \multicolumn{1}{l}{High} \\ \cmidrule(l){3-7} 
\multicolumn{1}{l}{} &
  \multicolumn{1}{l}{} &
  \multicolumn{1}{l}{ChestX-ray8~\cite{wang2017chestx}} &
  \multicolumn{1}{l}{108,948 images} &
  \multicolumn{1}{l}{8 labels per image} &
  \multicolumn{1}{l}{Disease detection} &
  \multicolumn{1}{l}{High} \\ \cmidrule(l){3-7} 
\multicolumn{1}{l}{} &
  \multicolumn{1}{l}{} &
  \multicolumn{1}{l}{BraTS~\cite{menze2014multimodal,bakas2017advancing,bakas2018identifying}} &
  \multicolumn{1}{l}{Varies annually} &
  \multicolumn{1}{l}{MRI data} &
  \multicolumn{1}{l}{Tumor segmentation} &
  \multicolumn{1}{l}{High} \\ \midrule
\multicolumn{1}{l}{\multirow{6}{*}{Multimodality}} &
  \multicolumn{1}{l}{\begin{tabular}[c]{@{}l@{}}Genomics, \\ Imaging\end{tabular}} &
  \multicolumn{1}{l}{TCGA~\cite{tomczak2015review}} &
  \multicolumn{1}{l}{Thousands of patients} &
  \multicolumn{1}{l}{Genomic and clinical data} &
  \multicolumn{1}{l}{Cancer research} &
  \multicolumn{1}{l}{High} \\ \cmidrule(l){2-7} 
\multicolumn{1}{l}{} &
  \multicolumn{1}{l}{\begin{tabular}[c]{@{}l@{}}Genomics, \\ Imaging, EHR\end{tabular}} &
  \multicolumn{1}{l}{UK Biobank~\cite{allen2014uk}} &
  \multicolumn{1}{l}{500,000 individuals} &
  \multicolumn{1}{l}{Various data types} &
  \multicolumn{1}{l}{Various tasks} &
  \multicolumn{1}{l}{Medium} \\ \cmidrule(l){2-7} 
\multicolumn{1}{l}{} &
  \multicolumn{1}{l}{\begin{tabular}[c]{@{}l@{}}Imaging, \\ Genomics, EHR\end{tabular}} &
  \multicolumn{1}{l}{ADNI~\cite{jack2008alzheimer}} &
  \multicolumn{1}{l}{Thousands of patients} &
  \multicolumn{1}{l}{MRI and clinical data} &
  \multicolumn{1}{l}{Alzheimer's research} &
  \multicolumn{1}{l}{High} \\ \cmidrule(l){2-7} 
\multicolumn{1}{l}{} &
  \multicolumn{1}{l}{\multirow{2}{*}{Imaging, Text}} &
  \multicolumn{1}{l}{ImageCLEFmed~\cite{GSB2016}} &
  \multicolumn{1}{l}{Varies annually} &
  \multicolumn{1}{l}{Various data types} &
  \multicolumn{1}{l}{Various tasks} &
  \multicolumn{1}{l}{Low} \\ \cmidrule(l){3-7} 
\multicolumn{1}{l}{} &
  \multicolumn{1}{l}{} &
  \multicolumn{1}{l}{Openi~\cite{demner2012design}} &
  \multicolumn{1}{l}{4.5 million images} &
  \multicolumn{1}{l}{Various data types} &
  \multicolumn{1}{l}{Various tasks} &
  \multicolumn{1}{l}{Low} \\ \cmidrule(l){2-7} 
\multicolumn{1}{l}{} &
  Various modalities &
  PhysioNet~\cite{PhysioNet} &
  Various datasets &
  Various data types &
  Various tasks &
  High \\ \bottomrule
\end{tabular}}
\footnotesize\textsuperscript{*}Popularity is determined by the citation count in Google Scholar as of 05/06/2023. It is categorized as Low ($\leq$200 citations), Medium ($>$200 and $<$1000 citations), and High ($\geq$1000 citations).
\end{table*}

\subsection{Behavioural data} 
Behavioural data plays a crucial role in smart healthcare by providing valuable insights into individuals' habits, lifestyles, and behaviours, which have a significant impact on their overall health and well-being. Behavioural data encompasses various aspects of human behaviour, including physical activity, sleep patterns, dietary habits, stress levels, social interactions, and adherence to medical treatments~\cite{feng2021multimodal}. By leveraging behavioural data, smart healthcare systems can monitor and analyze individuals' behaviours in real time, allowing healthcare providers to comprehensively understand their patients' daily routines and habits~\cite{zeadally2021harnessing}. This data can help identify patterns, trends, and deviations from normal behaviour, enabling early detection of potential health issues and the implementation of timely interventions.

Stress levels and emotional well-being can also be monitored through behavioural data, enabling healthcare providers to identify triggers and patterns that may impact individuals' mental health~\cite{woodward2020beyond}. This data has been used to guide the development of stress management techniques, relaxation strategies, and personalised interventions to support individuals in maintaining good mental health~\cite{soklaridis2020mental}. Furthermore, behavioural data can facilitate patient engagement and self-management. 

Through interactive platforms and feedback mechanisms, individuals can actively participate in monitoring their own behaviours, goal-setting, and receive personalised recommendations based on their provided data. This empowerment can lead to increased motivation and accountability in someone managing their health and well-being. However, the collection and analysis of behavioural data raises important ethical considerations, including privacy, data security, and informed consent. It is crucial to ensure that individuals' privacy is protected, their data is securely stored, and proper consent is obtained for data collection and usage.

\subsection{Multimodality Data}
Multimodality data fusion in smart healthcare involves integrating information from various sources, such as EHRs, medical imaging, wearable devices, genomic data, sensor data, environmental data, and behavioural data. By combining data from different modalities, healthcare professionals can gain a comprehensive understanding of a patient's health, leading to personalised care and informed decision-making. Each modality provides unique insights, and their fusion enhances the accuracy and completeness of the analysis. For example, in a fusion approach, EHRs provide historical medical records, medical imaging offers anatomical details, wearables capture real-time physiological data, genomics reveal genetic predispositions, sensors provide contextual information, and behavioural data reflect lifestyle choices. 

By integrating these modalities, healthcare professionals can uncover hidden patterns, correlations, and relationships that contribute toward techniques for optimising treatment strategies, predicting disease progression, identifying risk factors, and implementing preventive measures. Multimodality data fusion is a crucial step towards a holistic approach for advancing smart healthcare and improving patient outcomes.

\subsection{Datasets for Multimodal Fusion for Smart Healthcare}
An overview of multimodal datasets used in smart healthcare is presented in Table~\ref{tab:datasets}. It includes information on the modality, dataset name, number of instances, number of attributes, task, popularity, and reference count. The datasets cover various modalities such as EHRs, Genomics, Imaging, and Text. Examples of datasets include the eICU Collaborative Research Database, TCGA, UK Biobank, MRNet, RSNA Pneumonia Detection Challenge, MURA, and ChestX-ray8. These datasets are used for diagnosis, prognosis, cancer research, disease detection, and image segmentation~\cite{bhowal2022fuzzy}. The popularity of the datasets depends on the reference counts, and the table provides reference counts for further exploration.

\section{SOTA Techniques in Multimodal Fusion for Smart Healthcare}\label{approaches}
Multimodal medical data fusion involves combining information from multiple modalities such as medical imaging, genomic data, EHRs, wearable devices, and more. In this section, we explore state-of-the-art (SOTA) techniques for multimodal fusion across these multiple modalities within the context of smart healthcare.


\subsection{Feature selection} 
Feature selection focuses on identifying and selecting relevant features from raw data to transform them into meaningful information. In the context of multimodal medical data fusion for smart healthcare, recent studies have highlighted the importance of feature selection. Albahri et al.\cite{albahri2023systematic} emphasized its role in effective decision-making and improved patient care by identifying the most relevant features, reducing dimensionality, and enhancing the accuracy and interpretability of the fusion process. Similarly, Alghowinem et al.\cite{alghowinem2020interpretation} emphasized its impact on improving result interpretability. By systematically applying feature selection techniques, healthcare professionals can extract the most relevant information from the diverse data sources available, to facilitate a more comprehensive understanding of a patient's health conditions and support informed decision-making in personalized healthcare settings. Feature selection ensures efficient and effective data analysis and integration, ultimately enhancing the quality of data-driven insights and improving patient outcomes in smart healthcare environments. 

Before fusion, it is important to perform feature selection within each modality separately. This can be achieved using various techniques, such as statistical tests, information gain, correlation analysis, or ML algorithms~\cite{zhang2020emotion}. By selecting relevant features within each modality, noise and irrelevant information can be reduced, leading to improved fusion outcomes~\cite{zhang2020multi}. Certain modalities may contain more inherent noise or provide less relevant information compared to others. In such cases, modality-specific feature selection methods can be employed to identify the most informative features within each modality~\cite{zhou2019latent}. This can be done by leveraging domain knowledge, statistical analysis, or ML techniques tailored to the specific modality. After performing feature selection within each modality, the next step is to select features that are relevant across different modalities. Cross-modal feature selection methods aim to identify features that carry complementary information from multiple modalities~\cite{kim2021learning}. These methods can involve techniques such as correlation analysis, mutual information, or joint optimization algorithms~\cite{hoang2022multimodal}.

Feature selection and fusion should be performed in a coordinated manner to optimize the overall process. The selected features can be used as input for the fusion algorithm, which combines the information from different modalities~\cite{qiu2022multi}. This integration can be achieved through techniques such as early fusion, late fusion, or hybrid fusion approaches, depending on the nature of the data and the problem at hand~\cite{abdel2020new}. 

It is important to evaluate the performance of the feature selection and fusion methods using appropriate evaluation metrics. Evaluation can involve assessing classification accuracy, regression performance, clustering quality, or other domain-specific evaluation criteria~\cite{zhou2021evaluating}. Cross-validation or independent validation on separate datasets can help validate the effectiveness of the feature selection techniques~\cite{hao2020multi}. In healthcare applications, interpretability and explainability of the selected features and fusion results are crucial for building trust and understanding the decision-making process~\cite{yang2022unbox}. Various methods can be employed to enhance interpretability, such as feature importance ranking, visualization techniques, or rule extraction algorithms~\cite{zhang2021survey}.

\begin{figure*}
    \centering
    \includegraphics[width=\textwidth]{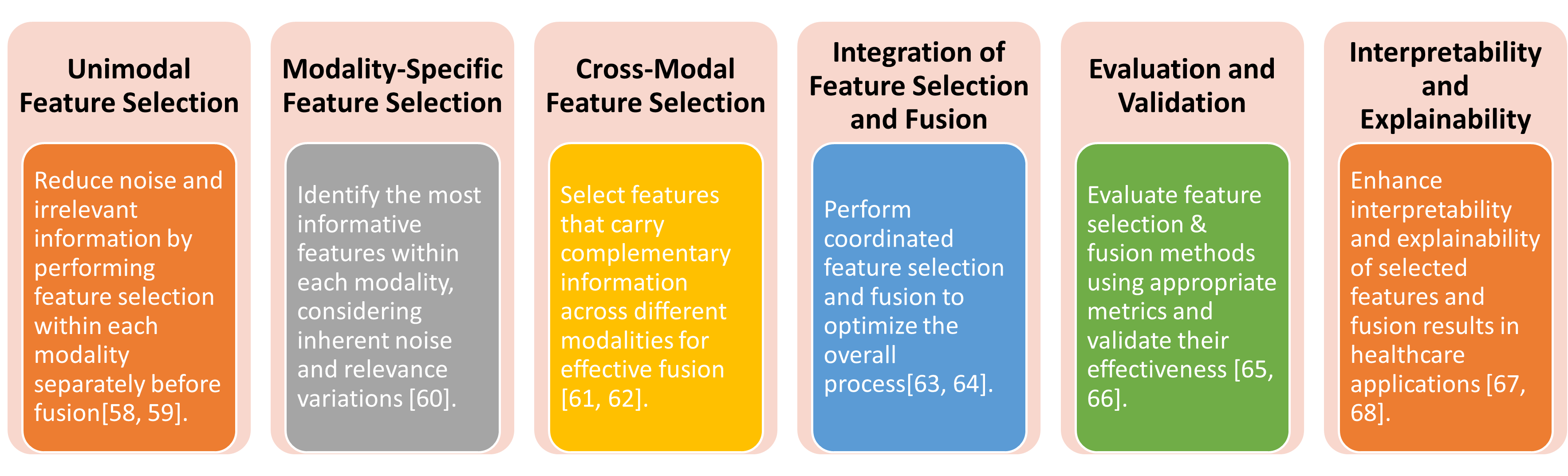}
    \caption{Multimodal Fusion - Feature Selection}
    \label{fig:feature_selection}
\end{figure*}

It's worth noting that the choice of feature selection methods may vary depending on the specific data characteristics, the fusion task, and the available computational resources. Additionally, the field of multimodal medical data fusion is an active area of research, and new techniques and algorithms are continuously being developed to address its challenges. The feature selection techniques of multimodal fusion are summarised in Fig.~\ref{fig:feature_selection}.

\subsection{Rule-based systems}
Rule-based systems operate to process and interpret information using predefined rules or logical statements. By employing these rules, these systems can make inferences and derive knowledge from the available data. The defined rules capture relationships and patterns, enabling decision-making based on the processed information. In the context of multimodal medical data fusion for smart healthcare, rule-based systems play a crucial role. They provide a structured approach to decision-making and knowledge representation~\cite{muhammad2021comprehensive}, offering a systematic framework for integrating information from various modalities. By employing a set of predefined rules, these systems can effectively process and integrate data from multiple sources, enabling informed decisions and providing valuable recommendations~\cite{hussain2021intelligent}. The utilization of rule-based systems in multimodal medical data fusion enhances the overall knowledge generation process, aiding in accurate diagnoses, personalized treatment plans, and improved healthcare outcomes.

Rules in a rule-based system are typically defined using an ``if-then'' format, meaning that each rule consists of an antecedent (if) and a consequent (then). The antecedent represents the conditions or criteria that need to be satisfied, while the consequent specifies the action or conclusion to be taken if the conditions are met~\cite{chen2021decision}. Rule-based systems can also contribute to feature selection in multimodal fusion. For example, rules can be designed to identify and select relevant features from different modalities based on their contribution to the decision-making process~\cite{yang2022unbox}. These rules can incorporate domain knowledge or statistical analysis to determine the importance of the features.
 
In multimodal medical data fusion, rules can be designed to accommodate multiple modalities. The antecedent of a rule can include conditions from different modalities, allowing the system to consider information from various sources simultaneously~\cite{mohd2022multi}. This integration can leverage the complementary nature of different modalities to enhance decision-making~\cite{yan2021richer}. 

Medical data often contains uncertainty and imprecision. Rule-based systems can leverage fuzzy logic, a mathematical framework that handles uncertainty, to model and reason with uncertain or imprecise data~\cite{amirkhani2018novel}. Fuzzy rules allow for more flexible decision-making by assigning degrees of membership to antecedents and consequents, capturing the inherent uncertainty in medical data fusion~\cite{geramian2019fuzzy}. In scenarios where multiple rules are applicable, conflicts may arise. Here, rule-based systems can employ strategies for rule prioritization and conflict resolution. These strategies determine the order in which rules are applied and resolve conflicts when multiple rules have conflicting conclusions~\cite{alharbi2020rule}. This ensures a systematic and consistent decision-making process.

Rule-based systems also offer transparency and interpretability by providing explicit rules that can be examined and understood by healthcare professionals~\cite{bahani2021accurate}. The rules provide explanations for the system's decisions, allowing users to understand the underlying reasoning process. This transparency is crucial in building trust and facilitating collaboration between clinicians and the decision-support system~\cite{antoniadi2021current}. Furthermore, rule-based systems enable the incorporation of expert knowledge into the decision-making process. Domain experts contribute their expertise by defining the rules that encapsulate their knowledge and clinical guidelines~\cite{hussain2021intelligent}. This allows the system to leverage the collective intelligence of healthcare professionals and enhance the accuracy and reliability of decision-making~\cite{wang2021knowledge}. 

By monitoring a system's performance and collecting feedback from its users, rule-based systems can be adapted or refined over time to improve decision-making~\cite{rundo2020recent}. This adaptive capability enables the system to evolve with new insights, changes in medical guidelines, or updates in the underlying data.

\begin{figure*}
    \centering
    \includegraphics[width=\textwidth]{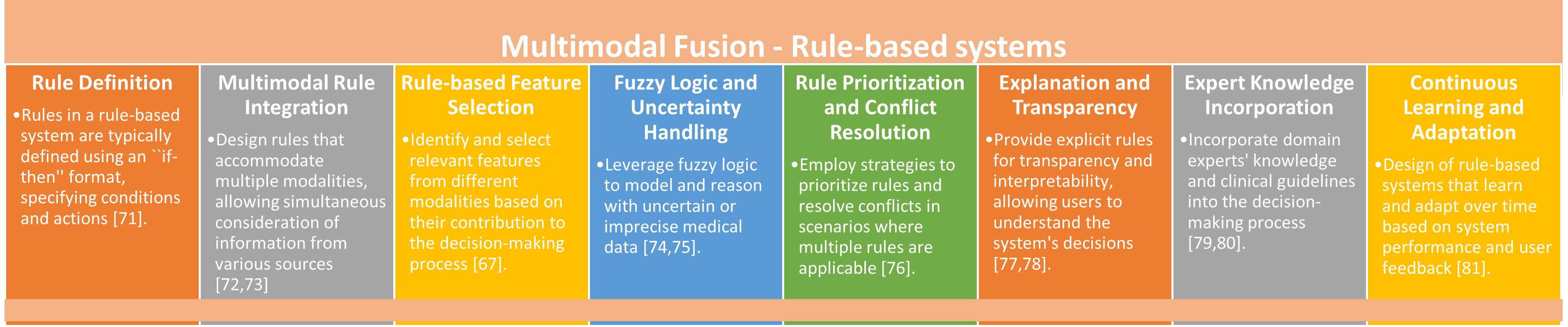}
    \caption{Multimodal Fusion - Rule-based systems}
    \label{fig:rule_based_sys}
\end{figure*}

Although rule-based systems provide a structured and interpretable framework for multimodal medical data fusion, it's important to consider the limitations of these approaches, such as the challenge of capturing complex relationships or interactions between modalities, and the potential for a large number of rules to manage. Hybrid approaches that combine rule-based systems with ML techniques can offer more flexibility and scalability in handling multimodal fusion tasks. The rule-based systems discussed in this subsection are outlined in Fig.~\ref{fig:rule_based_sys}.

\subsection{Machine Learning} 
Machine Learning (ML) encompasses the creation of algorithms capable of learning from data, enabling them to make predictions and informed decisions. By analyzing and processing vast amounts of data, ML algorithms can identify patterns, extract knowledge, and generate valuable insights to support decision-making processes. In the context of smart healthcare, ML techniques play a critical role in multimodal medical data fusion. These methods harness the capabilities of algorithms and statistical models to autonomously detect and understand patterns, relationships, and representations within diverse medical data sources. Through this automated learning process, ML contributes to overall knowledge generation within the healthcare sector, facilitating accurate diagnoses, personalized treatment plans, and improved patient outcomes.

Ensemble methods, such as Random Forests, gradient boosting, or AdaBoost, can be employed to combine the predictions or decisions of multiple ML models that have been trained on different modalities. Each modality can be processed independently using suitable algorithms, and their outputs can be fused using ensemble techniques to make a final decision~\cite{el2022automatic}. Ensemble learning helps leverage the diversity and complementary information present in different modalities~\cite{srinivasan2021wit}. Its adaptive weights combination approach works by assigning weights to different modalities based on their relevance or importance for the fusion task. The weights can be learned using various techniques, such as feature selection algorithms, statistical analysis, or ML models~\cite{yan2022emotion}. The modalities are then combined using weighted fusion strategies, such as weighted averaging or weighted voting, to generate a fused representation~\cite{de2021weighted}.

Bayesian networks provide a probabilistic graphical model framework for representing and reasoning about uncertainty in medical data fusion. Each modality can be treated as a node in the network, and the dependencies between modalities can be modelled using conditional probability distributions~\cite{gaebel2020modeling}. Bayesian networks allow for principled fusion of multimodal information, enabling probabilistic inference and decision-making~\cite{chen2020multimodality}. 

Multiple Kernel Learning (MKL) is a technique that combines multiple kernels, which capture different types of information or relationships, into a unified representation. Each modality can be represented using a separate kernel, and MKL methods can learn the optimal combination of these kernels to maximize the performance of the fusion task~\cite{cao2018ℓ2}. MKL allows for flexible and effective integration of information from different modalities. Feature-level fusion techniques combine features extracted from different modalities to create a unified feature representation. This can involve techniques like concatenation, feature stacking, or feature selection, based on relevance or mutual information~\cite{sharma2022comprehensive}. The fused features can then be used as input for traditional ML algorithms, such as support vector machines (SVM), logistic regression, or k-nearest neighbours (k-NN), to perform classification, regression, or clustering tasks~\cite{lopez2020clinical}.

Canonical Correlation Analysis(CCA) is a statistical technique that aims to find linear transformations of multiple modalities to maximize their correlation. It identifies common underlying factors that explain the correlations between modalities~\cite{liu2018identifying}. The fused representation can then be used as input for subsequent ML algorithms~\cite{zhang2020multimodal}. Manifold learning techniques, such as t-SNE (t-Distributed Stochastic Neighbor Embedding) or Isomap, can be used to map multimodal data into a lower-dimensional space, while preserving the underlying structures and relationships~\cite{anowar2021conceptual}. By projecting the multimodal data onto a common latent space, these techniques facilitate the fusion of modalities and enable visualization and analysis of the fused data.

\begin{figure*}[h]
    \centering
    \includegraphics[width=\textwidth]{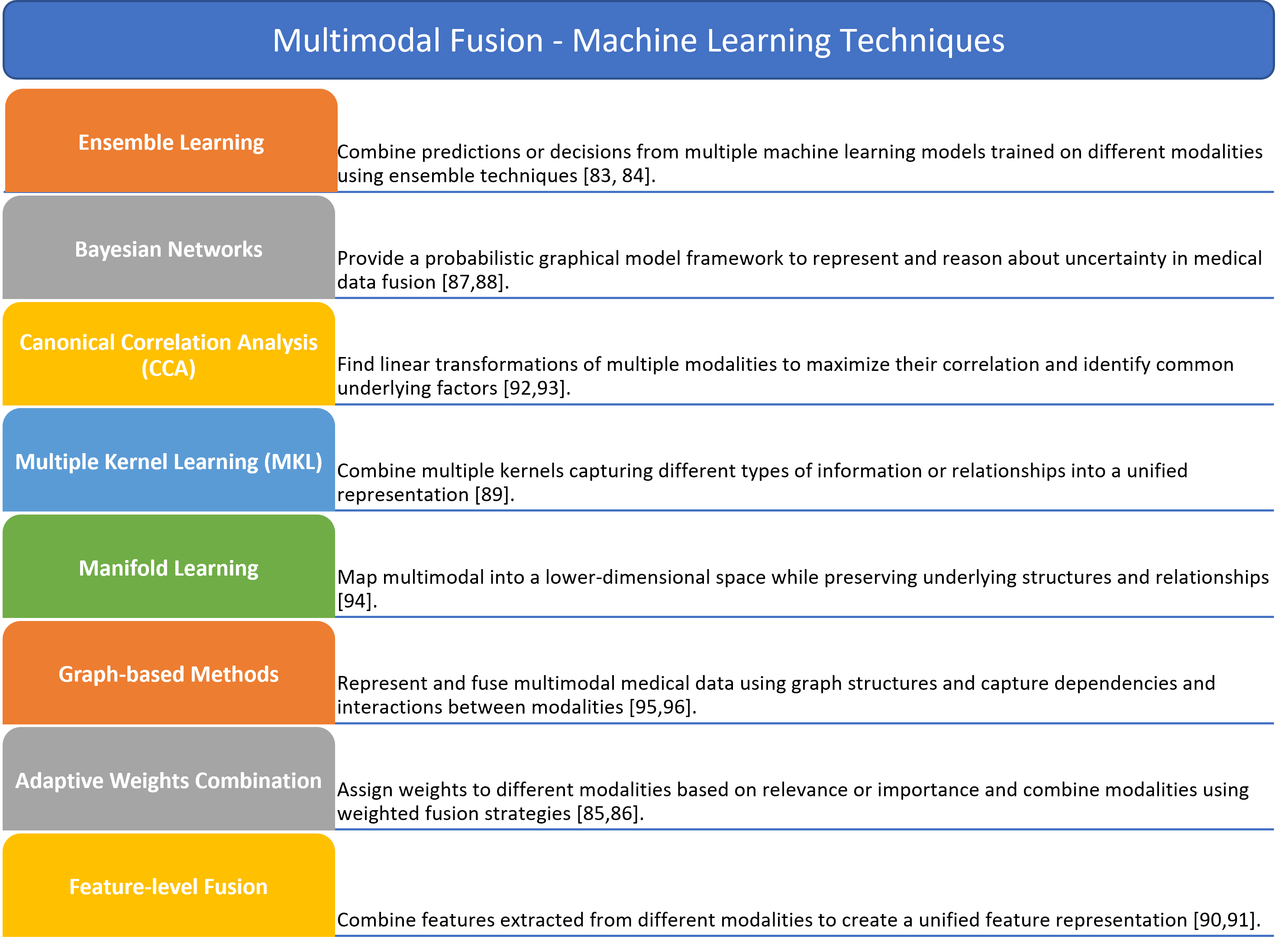}
    \caption{Multimodal Fusion - Machine Learning Techniques}
    \label{fig:ML_techniques}
\end{figure*}

Graph-based methods offer a framework for representing and fusing multimodal medical data using graph structures. Each modality can be represented as nodes, and edges can be defined based on the relationships or correlations between modalities~\cite{zheng2022multi}. Graph-based algorithms, such as graph convolutional networks (GCNs) or graph regularized non-negative matrix factorization (GNMF), can then be applied to capture the dependencies and interactions between modalities~\cite{yan2018spatial}.

ML techniques, even without deep learning, can still be effective in multimodal medical data fusion for smart healthcare. It's important to note that the selection of specific ML techniques for multimodal medical data fusion depends on the nature of the data, the fusion task, and the available computational resources. Careful consideration should be given to feature selection, normalization, and data preprocessing steps to ensure optimal fusion performance. Additionally, model evaluation and validation using appropriate metrics and cross-validation techniques are crucial to assess the effectiveness of the fusion approach in smart healthcare applications. In Fig.~\ref{fig:ML_techniques}, the ML techniques in multimodal fusion are presented.

\subsection{Deep learning} 
Deep learning, as a subset of ML, plays a crucial role at the knowledge level of the DIKW framework. It specializes in training neural networks with multiple layers to extract intricate features and representations from data. By processing vast amounts of data, deep learning models can learn complex patterns and generate valuable knowledge for decision-making purposes. In the context of smart healthcare, deep learning has emerged as a potent approach for multimodal medical data fusion. Its unique capability to automatically learn hierarchical representations from diverse and complex medical modalities makes it particularly well-suited for integrating and extracting meaningful information. With its ability to handle diverse data types and capture intricate relationships, deep learning contributes to the overall knowledge-generation process in healthcare, enabling more accurate diagnoses, personalized treatment plans, and improved patient outcomes. Here are some key aspects of deep learning in multimodal medical data fusion.

Deep learning architectures, such as convolutional neural networks (CNNs), recurrent neural networks (RNNs), or transformers, can learn representations directly from multimodal medical data~\cite{hugle2021dynamic}. By jointly processing multiple modalities, these models capture both local and global dependencies, enabling the extraction of rich and informative features. This facilitates the fusion of modalities at various levels, ranging from low-level pixel or waveform data to high-level semantic representations~\cite{elboushaki2020multid}. Deep learning models with recurrent or temporal components, such as RNNs or long short-term memory (LSTM) networks, can handle sequential or temporal aspects of multimodal medical data~\cite{rashid2019times}. These models can capture temporal dependencies, changes over time, or dynamic patterns across modalities. This is particularly relevant for applications such as physiological signal analysis, time-series data fusion, or modelling disease progression~\cite{bahador2021multimodal}.

Deep learning models that have been pre-trained on large-scale datasets, such as ImageNet or natural language corpora, can be leveraged for multimodal medical data fusion~\cite{wang2023large}. Transfer learning techniques allow the transfer of knowledge from pretraining to the medical domain, enabling the models to learn relevant representations from limited medical data. This approach can boost performance, especially when multimodal medical datasets are small or resource-intensive to collect~\cite{ayana2021transfer}. Attention mechanisms in deep learning models provide a mechanism for focusing on salient regions or modalities within the input data. They learn to allocate attention to the most relevant features or modalities, enhancing the fusion process~\cite{de2022attention}. Attention mechanisms can be employed within CNNs, RNNs, or transformer architectures to selectively combine or weigh the contributions of different modalities based on their importance for the task at hand~\cite{niu2021review}.
    
Generative models, such as generative adversarial networks (GANs) or variational autoencoders (VAEs), can be used for multimodal fusion. These models learn to generate new samples from the joint distribution of multiple modalities, capturing their underlying correlations~\cite{shi2019variational}. Generative models can aid in data augmentation, missing data imputation, or synthesis of multimodal data, facilitating improved training and fusion outcomes~\cite{du2021multimodal}. Deep fusion architectures combine multiple modalities at different stages of the network, allowing for the explicit integration of multimodal information. For example, early fusion involves combining modalities at the input level, while late fusion integrates modalities at higher layers or during decision-making~\cite{joze2020mmtm}. Hybrid fusion approaches leverage both early and late fusion strategies to capture complementary information effectively. Deep fusion architectures can enhance the performance and robustness of multimodal fusion tasks~\cite{zhang2021deep}.
    
Deep learning models will benefit from the integration of clinical knowledge, domain expertise, or prior medical information. Architectures that incorporate domain-specific constraints, expert rules, or Bayesian priors can enhance the fusion process and align the models with established medical knowledge~\cite{carvalho2021integrating}. Integrating clinical knowledge helps improve the interpretability, reliability, and acceptance of deep learning models in smart healthcare settings~\cite{jin2022explainable}. 

Deep learning models, although powerful, can be challenging to interpret. However, techniques such as attention visualization, saliency mapping, or gradient-based methods can provide insights into the model's decision-making process~\cite{sevastjanova2018going}. Interpretable deep learning architectures, such as CNNs with structured receptive fields or interpretable RNN variants, are also being explored to enhance transparency and explainability in multimodal medical data fusion~\cite{boehm2022harnessing}.

\begin{figure}[h]
    \centering
    \includegraphics[width=\columnwidth]{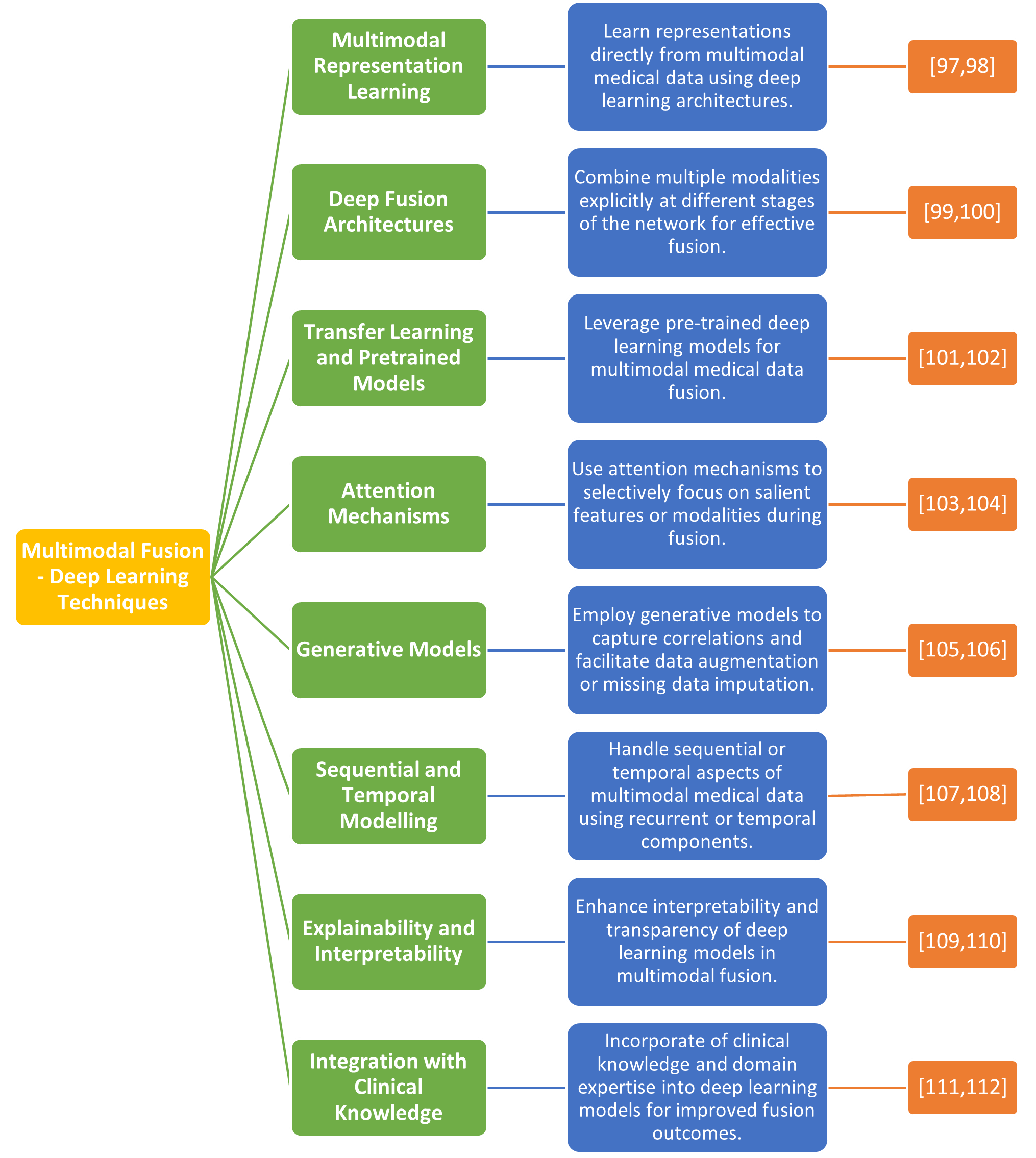}
    \caption{Multimodal Fusion - Deep Learning Techniques}
    \label{fig:deep_learn_tech}
\end{figure}

Deep learning techniques offer promising avenues for multimodal medical data fusion, but challenges such as the need for large labelled datasets, interpretability, and generalization to new patient populations must be addressed. Collaboration between deep learning researchers, healthcare professionals, and data scientists is crucial to developing effective and reliable deep learning approaches for multimodal medical data fusion in smart healthcare. The deep learning techniques for multimodal fusion are outlined in Fig.~\ref{fig:deep_learn_tech}

\subsection{Natural Language Processing} 
NLP plays a vital role in transforming textual data into structured information, uncovering insights, and facilitating informed decision-making. In the context of multimodal medical data fusion for smart healthcare, NLP is particularly valuable for processing textual information from clinical notes, reports, and records. It extracts relevant details, identifies relationships, and discovers hidden patterns within the text. By incorporating NLP into the data fusion process, healthcare professionals can gain a comprehensive understanding of patients' health, improve diagnosis accuracy, and enhance personalized treatment planning. NLP is a powerful tool for integrating text-based information with other data modalities, enabling a holistic approach to healthcare decision-making.

NLP techniques are employed to process and extract meaningful information from unstructured textual data. Tasks such as tokenization, sentence segmentation, part-of-speech tagging, named entity recognition, and syntactic parsing help structure and analyze clinical text~\cite{shaik2022review}. NLP enables the extraction of relevant concepts, medical terms, and relationships from textual data, facilitating their integration with other modalities~\cite{zeng2018natural}. NLP techniques can extract structured information from clinical narratives, such as diagnoses, medications, procedures, and patient demographics. Named entity recognition and relationship extraction algorithms identify and classify relevant entities and their associations, contributing to the fusion of textual information with other modalities. This extracted information can be used for decision support, clinical coding, or cohort identification~\cite{bhatia2019comprehend}. NLP techniques, including semantic parsing, semantic role labeling, and medical concept normalization, each of which enable the understanding of clinical text in a structured manner~\cite{demner2021natural}. This facilitates the extraction of clinical concepts, relations, and contextual knowledge, which can be fused with other modalities for comprehensive analysis, decision support, or knowledge discovery~\cite{petrova2019towards}.

NLP models can be trained to classify clinical text into various categories, such as disease categories, severity levels, or treatment options~\cite{pham2020constructing,tang2019progress}. Sentiment analysis techniques can also assess the sentiment or opinion expressed in patient feedback, social media data, or clinical notes. Text classification and sentiment analysis provide valuable insights, and can be integrated with other modalities for a comprehensive understanding of the patient's condition~\cite{tao2020mining, chintalapudi2021text}. NLP can also bridge the gap between textual and visual modalities in medical data fusion. By analyzing textual descriptions or radiology reports, NLP techniques can extract relevant information about anatomical locations, findings, or abnormalities~\cite{bozkurt2019automated}. This information can be linked to corresponding images or visual data, enabling the fusion of text and image modalities for improved diagnosis, treatment planning, or image interpretation~\cite{pei2023review}.

NLP methods aid in identifying and monitoring adverse events by analyzing textual data sources such as EHRs, patient complaints, or pharmacovigilance reports~\cite{wang2018detecting}. Sentiment analysis, information extraction, and text mining techniques can automatically detect and categorize adverse events \cite{guinazu2020employing}, enabling timely interventions and enhancing patient safety~\cite{choudhury2020role}. NLP techniques contribute to patient risk assessment by analyzing clinical narratives and extracting relevant information related to patient history, comorbidities, and lifestyle factors~\cite{le2021machine}. By integrating this textual information with other patient data, such as vital signs, imaging results, or genetic information, multimodal medical data fusion can provide a comprehensive risk assessment strategy for personalized healthcare interventions and preventive measures~\cite{lipkova2022artificial}. 

NLP techniques support the development of clinical decision support systems by extracting relevant clinical knowledge from medical literature, clinical guidelines, or research articles~\cite{hiremath2022enhancing,spasic2020clinical}.

\begin{figure}[h]
    \centering
    \includegraphics[width=\columnwidth]{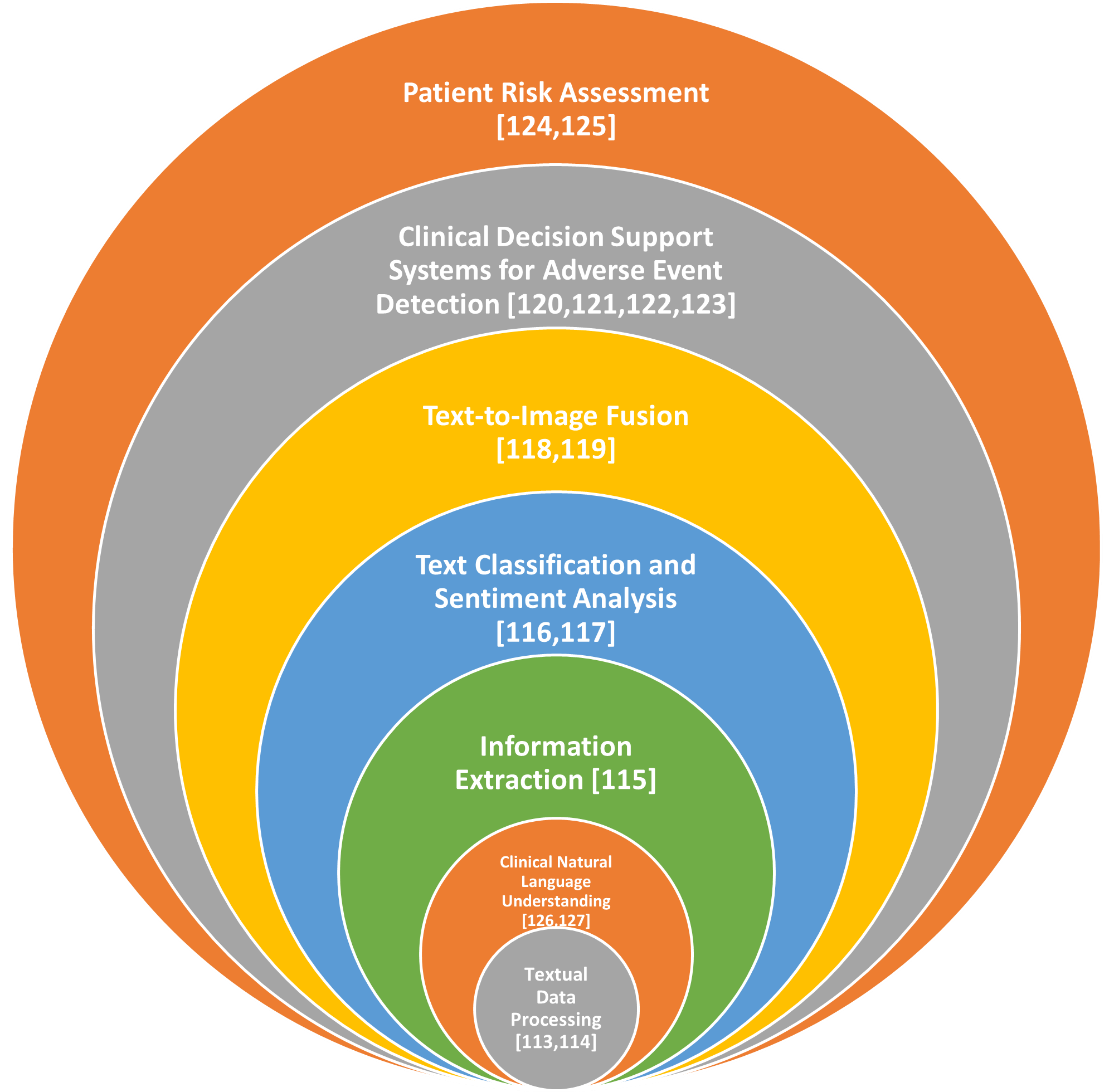}
    \caption{Multimodal Fusion - NLP techniques}
    \label{fig:nlp_tech}
\end{figure}

NLP techniques provide valuable capabilities in extracting, processing, and integrating textual information within multimodal medical data fusion. They improve the comprehension of clinical text, enable the incorporation of clinical knowledge, and facilitate comprehensive and effective analysis in smart healthcare applications. Fig.~\ref{fig:nlp_tech} presents the NLP techniques that can be adopted for multimodal fusion.

    \begin{figure*}
        \centering
        \includegraphics[width=\textwidth]{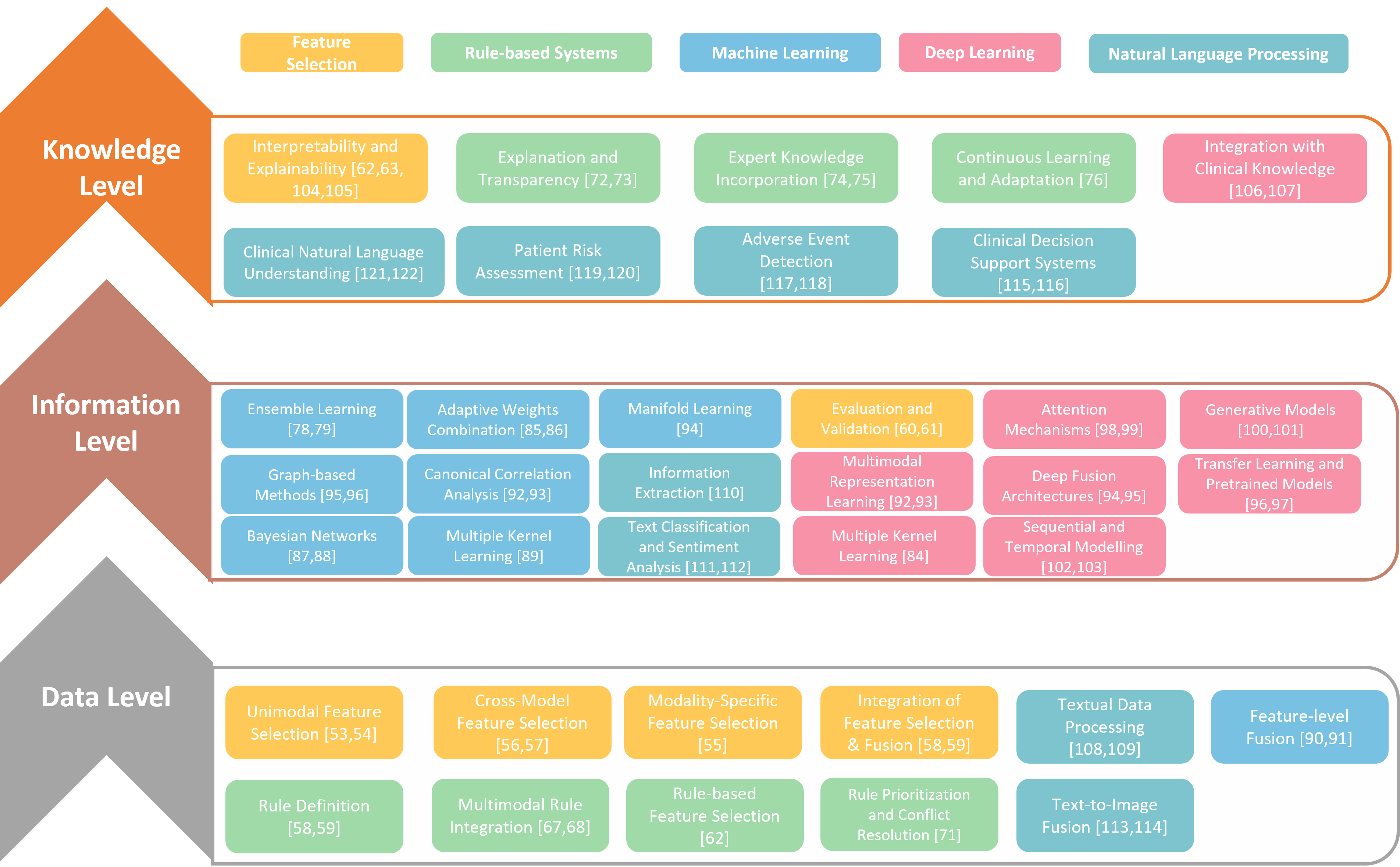}
        \caption{Taxonomy of SOTA techniques in Multimodal Fusion for Smart Healthcare}
        \label{fig:taxonomy}
    \end{figure*}

\subsection{Taxonomy of Approaches in Multimodal Fusion}
The taxonomy of approaches in multimodal fusion for smart healthcare encompasses feature selection, rule-based systems, ML, deep learning, and NLP, as shown in Fig.~\ref{fig:taxonomy}. These techniques play a crucial role in integrating and analyzing diverse data modalities to extract valuable insights and support informed decision-making in healthcare applications. 
\begin{itemize}
    \item Feature selection focuses on identifying relevant features to create a concise representation of the data.
    \item  Rule-based systems utilize predefined rules to process and combine data from multiple modalities.
    \item ML leverages patterns and relationships in the data to predict, classify, or cluster information from different modalities.
    \item Deep learning employs neural networks to automatically learn hierarchical representations and capture complex relationships.
    \item NLP techniques process and analyze textual information, enhancing the understanding of clinical data and facilitating its integration with other modalities.
\end{itemize}

By leveraging these approaches, researchers and healthcare professionals can gain a deeper understanding of patient health, enable personalized care, and make more informed decisions in intelligent healthcare systems, ultimately advancing patient care and well-being. In Fig.~\ref{fig:taxonomy}, we provide a comprehensive taxonomy of multimodal fusion methods relevant to smart healthcare. The Fig.~\ref{fig:taxonomy} categorizes SOTA techniques, including feature selection, rule-based systems, machine learning, deep learning, and NLP, and aligns them with the different levels of the DIKW conceptual model.

\section{Challenges in adopting Multimodal Fusion}\label{challenges}

There are numerous challenges in adopting multimodal data fusion approaches that influence different stages and aspects of the DIKW framework. These challenges include issues such as data quality and interoperability, privacy and security, data processing and analysis, clinical integration and adoption, ethical considerations, and interpretation of results, all of which impact the transformation of data into meaningful information, knowledge, and wisdom in healthcare.

\subsection{Data quality and interoperability}
Data quality and interoperability present significant challenges in the context of multimodal fusion for smart healthcare within the DIKW framework. The integration of data from diverse sources and ensuring its quality and compatibility across different healthcare systems and modalities can be complex and time-consuming~\cite{perez2020future}. Insufficient data quality and a lack of interoperability can result in inaccurate analysis and hinder the effectiveness of the data fusion process~\cite{sun2022self}.

Addressing these challenges necessitates the development and adoption of data standards and protocols. Standardizing data formats, such as HL7 for EHRs or DICOM for medical imaging, facilitates seamless integration and data exchange across healthcare systems and modalities~\cite{reegu2023blockchain}. These standards establish a shared language for data representation, simplifying the processing and integration of data from various sources~\cite{lyketsos2022standardizing}. Establishing interoperability frameworks plays a vital role in promoting smooth data sharing and exchange among different healthcare systems and modalities~\cite{diraco2023review}. These frameworks provide guidelines and best practices for data integration, harmonization, and transmission, since they define the protocols, data models, and communication standards that enable efficient interoperability across diverse the data sources. Adhering to interoperability frameworks enhances data compatibility and coherence, thereby facilitating effective multimodal fusion~\cite{mwangi2022assessing}. 

Advancements in ML techniques offer promising avenues for addressing data integration and interoperability challenges~\cite{kor2023investigation}. ML algorithms can learn patterns and relationships in data obtained from different sources, facilitating automated data mapping, harmonization, and integration. Leveraging ML enables organizations to streamline the data fusion process, improving efficiency and accuracy in multimodal integration~\cite{tao2022multi}.

\subsection{Privacy and security}
Privacy and security pose significant challenges in the integration of sensitive patient data from multiple sources within the DIKW framework. Protecting patient privacy and ensuring data security are paramount concerns in healthcare, particularly when dealing with sensitive health information~\cite{paul2023digitization}. Robust privacy and security measures are necessary to safeguard patient confidentiality and prevent unauthorized access or data breaches in the integration of multimodal medical data.

To address these challenges, implementing data encryption techniques is essential to protect patient data during transmission and storage~\cite{yasser2021robust}. Encryption converts data into an unreadable form, ensuring that only authorized individuals with decryption keys can access and interpret the data~\cite{regade2022survey}. Secure storage methods, such as secure servers or cloud platforms with robust access controls, play a crucial role in safeguarding patient data from unauthorized access, loss, or theft~\cite{al2019ehealth}. In addition, adopting privacy-preserving techniques is crucial to protect patient privacy during data fusion. Differential privacy, for example, adds noise to aggregated data to prevent individual identification while preserving the utility of the fused data~\cite{mohammed2020internet, kim2021learning}. Secure multiparty computation (SMC) techniques allow collaboration on data fusion without revealing individual-level data to any party involved, ensuring privacy during the fusion process~\cite{hathaliya2022adversarial}. The continuous monitoring and auditing of data access and usage are vital in detecting and preventing unauthorized activities and potential data breaches~\cite{neto2021case}. Robust auditing mechanisms and log analysis techniques enable organizations to track and investigate any suspicious or anomalous access patterns or data breaches~\cite{kumar2019cloud}. Real-time monitoring systems can provide alerts and notifications in case of any unauthorized access attempts or potential security incidents~\cite{kebande2021real}. By implementing these privacy and security measures, healthcare organizations can ensure the protection of patient data, maintain privacy during data fusion, and mitigate the risks associated with unauthorized access or data breaches.

\subsection{Data processing and analysis}
Data processing and analysis play a critical role in the DIKW framework, particularly in multimodal medical data fusion. Challenges arise in handling large volumes of data, scalability of data processing, and extracting actionable insights from the fused data~\cite{bokade2021cross}. To address these challenges, ML algorithms and AI techniques are leveraged to enable efficient processing and analysis of multimodal medical data~\cite{yang2022unbox, flores2021leveraging}. Supervised and unsupervised learning algorithms are utilized for classification, prediction, and pattern discovery~\cite{swathy2022comparative}. Deep learning models, such as CNNs and RNNs, are applied to tasks involving medical imaging and sequential data analysis~\cite{banerjee2019comparative}. Reinforcement learning techniques optimize treatment plans and interventions based on patient outcomes~\cite{coronato2020reinforcement}. 

Collaboration between clinicians and data scientists are crucial in developing effective data processing and analysis solutions that align with clinical needs~\cite{wang2019human, sarker2021data}. Integration of multimodal medical data fusion into Clinical Decision Support Systems (CDSS) enhances clinical decision-making and improves patient outcomes~\cite{steyaert2023multimodal, wang2021brilliant}. Scalable data processing techniques, including distributed computing frameworks and cloud computing platforms, handle large-scale datasets~\cite{nazari2019bigdata, kaur2020fog}. Real-time data analytics enable immediate insights and proactive interventions~\cite{dwivedi2022potential}. Effective visualization techniques can aid in interpreting and communicating analysis results~\cite{qi2021enabling}. By addressing these challenges and utilizing advanced data processing and analysis techniques, healthcare organizations can unlock the full potential of multimodal medical data fusion within the DIKW framework.

\subsection{Clinical integration and adoption}
Clinical integration and adoption present significant challenges in the successful implementation of multimodal fusion within the DIKW framework in healthcare~\cite{yang2022unbox}. To address these challenges, involving clinicians and other stakeholders in the development and implementation process is crucial for both ensuring successful adoption and maximizing the impact of multimodal fusion in clinical practice~\cite{maddikunta2022industry, vakil2021survey}. Their input and feedback contributes toward designing technologies that align with clinical workflows and meet end-users needs~\cite{you2023federated}.

The design of user-friendly interfaces and intuitive workflows is essential to facilitate the integration of multimodal fusion into clinical practice. Applying user-centered design principles and conducting usability testing can identify and address usability issues, enhancing user satisfaction and adoption rates~\cite{dabliz2021usability, rundo2020recent}. Integrating multimodal fusion technologies with CDSS can enhance clinical decision-making processes by providing real-time recommendations and alerts based on the fused data~\cite{steyaert2023multimodal, limketkai2021age}. Embedding these technologies into familiar clinical tools streamlines the integration process and facilitates adoption. To ensure successful adoption, it is crucial to provide adequate training and education to healthcare professionals~\cite{chen2023information, o2020paediatric}. Training programs should focus not only on the technical aspects, but also on the clinical relevance and potential impacts on patient care. Continual education and support help healthcare professionals remain proficient in using the technologies and stay updated on advancements.

\subsection{Ethical considerations}
Ethical considerations are of utmost importance in the context of multimodal medical data fusion within the DIKW framework. Ensuring patient privacy, autonomy, and fairness is essential in utilizing patient data ethically~\cite{holzinger2022information}. Obtaining informed consent from patients is a fundamental ethical requirement in multimodal medical data fusion. Patients should be fully informed about the purpose, risks, and benefits of data fusion, and consent processes should be transparent and understandable~\cite{baum2023data}. Clear mechanisms for patients to withdraw their consent should also be provided. 

Defining data ownership and governance policies is crucial. Healthcare organizations should establish guidelines to determine data ownership, usage, and access~\cite{van2021governance}. Transparent governance mechanisms, such as data access committees, should oversee the ethical use of patient data and compliance with privacy regulations. Respecting patient privacy and ensuring data confidentiality is paramount. Robust security measures, including encryption and access controls, should be implemented~\cite{thapa2021precision}. Compliance with privacy regulations like HIPAA or GDPR should be ensured. 

Addressing potential biases is essential in multimodal fusion. Efforts should be made to mitigate biases through rigorous data collection processes, algorithmic fairness assessments, and diverse representation in data and development teams~\cite{gaw2022multimodal}. Regular monitoring and auditing can help identify and address biases. Ethical frameworks and guidelines should be developed and followed. These frameworks should outline principles and best practices for ethical data collection, fusion, analysis, and decision-making~\cite{mokander2021ethics}. Guidelines should cover areas such as data privacy, informed consent, fairness, transparency, and accountability. Engaging the public and stakeholders in discussions about ethical considerations is crucial. Open communication channels should be established to foster trust and incorporate patient perspectives into decision-making processes~\cite{belgodere2023patient}. By addressing ethical considerations, healthcare organizations can ensure the responsible and ethical use of patient data in multimodal medical data fusion, promoting patient privacy, fairness, and trust within the DIKW framework.

\subsection{Interpretation of results}
Interpreting the results of multimodal medical data fusion within the DIKW framework can be challenging due to the complexity of integrating multiple modalities and the large volumes of generated data. Effective interpretation is crucial for extracting meaningful insights and making actionable decisions in clinical settings~\cite{ali2023explainable}.

To facilitate interpretation, the development of visual analytics tools and techniques is essential. Interactive visualizations, such as heatmaps, scatter plots, or network diagrams, can assist clinicians in identifying patterns, correlations, and outliers in the fused data~\cite{rostamzadeh2021visual}. These visualizations should provide intuitive representations and enable interactive exploration at different levels of detail~\cite{hollt2019focus+}. Enhancing the interpretability of fusion models and algorithms is another important aspect. Techniques like explainable AI, interpretable machine learning, or rule-based systems can provide transparent explanations for the fusion process and decision-making~\cite{arrieta2020explainable}. 

Understanding how the fusion models arrive at certain conclusions helps clinicians trust the results and make informed decisions based on the interpretation of the fused data. Clinical validation studies are crucial for assessing the clinical utility and effectiveness of the interpretation results. Real-world evaluation with healthcare professionals provides insights into the practical applicability of the interpretation and helps refine the techniques to ensure meaningful and actionable results are aligned with clinical practice. Involving domain experts, such as clinicians or medical researchers, in the interpretation process is vital. They bring valuable insights into the clinical relevance of the fused data and help interpret the results within the context of patient care~\cite{Holzinger2022}. 

Collaboration between data scientists and domain experts fosters mutual understanding and enables the development of interpretation techniques that meet the specific needs of healthcare professionals~\cite{Mao2019}. Incorporating clinical guidelines and contextual information into the interpretation of fused data is crucial. Considering patient-specific factors, such as demographics, medical history, or treatment guidelines, helps provide personalized interpretations and recommendations~\cite{Mller2021}. Aligning the interpretation with existing clinical knowledge and guidelines ensures clinically meaningful and actionable results for healthcare providers.






\begin{figure*}
    \centering
    \includegraphics[width=\textwidth]{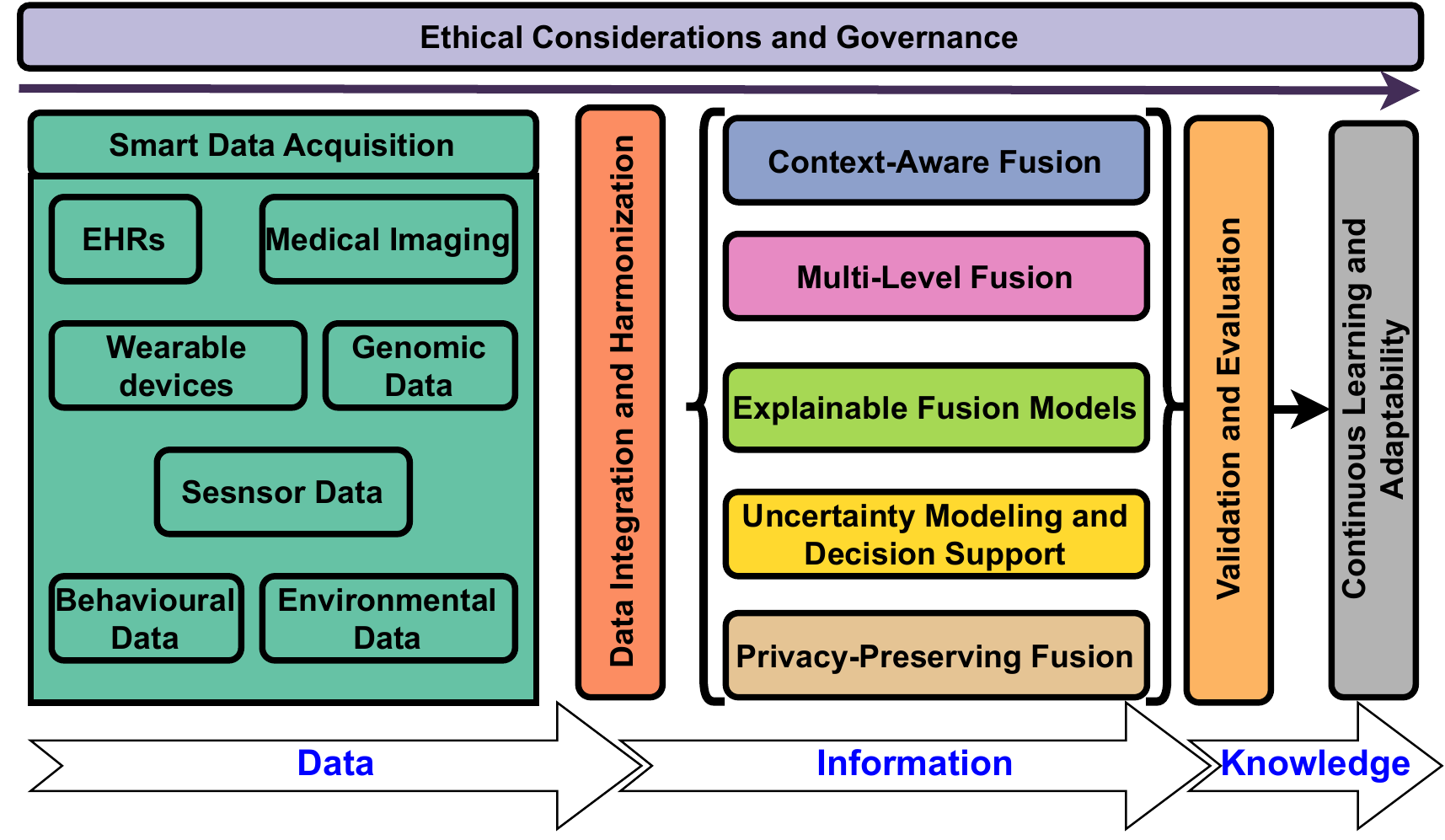}
    \caption{Generic Framework of Multimodal Fusion for Smart Healthcare}
    \label{fig:framework}
\end{figure*}

\section{DIKW Fusion Framework with Multimodality}\label{generic}

Based on the data representation modalities and multimodal fusion approaches, we present a universal framework that can be applied to diverse applications in Fig.~\ref{fig:framework}. In the framework for multimodal fusion in smart healthcare, several components are identified that facilitate the progression towards wisdom. At the data level, strategies for efficient data acquisition are employed, followed by data integration and harmonization processes to create a unified dataset. 

Moving to the information level, context-aware fusion incorporates contextual information to enhance the fusion process, while multi-level fusion techniques capture complex relationships and patterns. Explainable fusion models provide transparency and trust, and uncertainty modeling supports decision-making based on fused data. Privacy-preserving fusion techniques ensure responsible data handling, and validation and evaluation methods assess the performance of the fusion framework. At the knowledge level, continuous learning and adaptability mechanisms enable the framework to stay up-to-date, while ethical considerations and governance frameworks address ethical issues in healthcare fusion.

\begin{figure*}
    \centering
    \includegraphics[width=\textwidth]{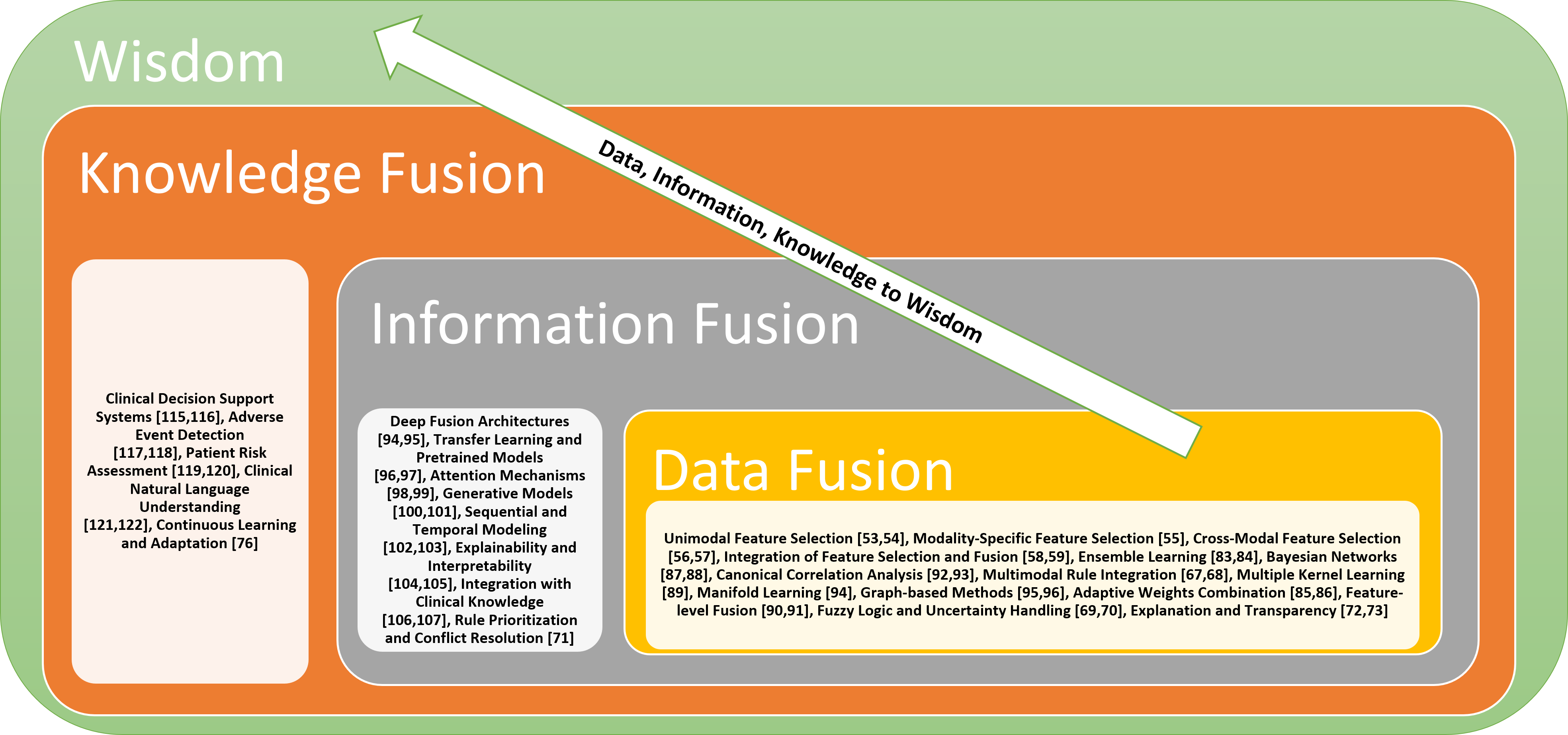}
    \caption{DIKW Conceptual Model Journey}
    \label{fig:data_to_wisdom_journey}
\end{figure*}

In the landscape of multimodal data fusion for healthcare applications, the journey toward wisdom can be conceptualized as a hierarchical framework consisting of four integral stages: data fusion, information fusion, knowledge fusion, and ultimately, wisdom, as shown in Fig.~\ref{fig:data_to_wisdom_journey}. In the first stage, data fusion, we employ techniques such as feature selection, ensemble learning, and graph-based methods to combine and select the most relevant features from diverse data sources~\cite{pham2022graph}. This crucial step forms a solid groundwork for effectively integrating and processing multimodal data.

Moving on to the second stage, information fusion, we delve deeper into the data by utilizing advanced techniques such as deep fusion architectures, transfer learning, attention mechanisms, and sequential modeling. These sophisticated approaches enable us to uncover intricate relationships and patterns across modalities, providing a more profound understanding of the data at hand. Additionally, we employ explainability and interpretability techniques to gain valuable insights into the decision-making process of the fusion models.

In the final stage, knowledge fusion, we take integration to the next level by incorporating clinical knowledge and domain expertise. Here, techniques like CDSS, adverse event detection, patient risk assessment, and clinical natural language understanding come into play. By leveraging this wealth of clinical knowledge, we can extract actionable insights and make informed decisions in the healthcare domain.

By following this progressive journey from data fusion to information fusion and knowledge fusion, we empower ourselves to enhance our understanding and analysis of multimodal data. This, in turn, contributes to the development of wisdom in the field of multimodal data fusion, enabling us to make more impactful advancements in smart healthcare and its applications.

\section{Future Directions of DIKW Fusion in Smart Healthcare}\label{future}
The field of multimodal medical data fusion for smart healthcare is expected to evolve in line with the 4Ps of healthcare – Predictive, Preventive, Personalized, and Participatory~\cite{collatuzzo2022application, ruiz2022artificial}. Predictive healthcare data fusion aims to anticipate health events and outcomes by combining data from various sources, while Preventive data fusion focuses on identifying risk factors and promoting healthy behaviors. Personalized fusion caters to individual-specific data for customized care, and Participatory fusion involves patients and stakeholders in the data fusion process, enhancing transparency and trust. The progress towards these goals forms the crux of our future research.

\subsection{Predictive Healthcare}
In the context of the DIKW framework and the generic framework discussed, the ``Predictive'' component of multimodal fusion in smart healthcare focuses on utilizing diverse data sources to anticipate health events and outcomes, thereby enabling proactive interventions and personalized healthcare strategies~\cite{collatuzzo2022application}. Within the DIKW framework, the Predictive component involves leveraging multimodal fusion to generate predictive models for assessing disease likelihoods~\cite{boehm2022harnessing}. By integrating and analyzing data from various modalities, such as genomics, medical imaging, and clinical records, healthcare professionals can identify early indicators or risk factors~\cite{esteva2019guide}. ML algorithms play a key role in analyzing the combined data and uncovering patterns indicative of disease risk~\cite{myszczynska2020applications}.

In the generic framework, the Predictive component of multimodal fusion aims to identify potential risk factors or biomarkers for disease prediction~\cite{lipkova2022artificial}. Healthcare professionals gain insights into genetic predispositions, imaging abnormalities, and the clinical context~\cite{zhang2020advances}. ML algorithms enable the development of predictive models by leveraging the combined data and considering variables such as genetic markers, imaging features, clinical parameters, lifestyle factors, and environmental exposures~\cite{bi2021novel, vale2021long}. 

By applying predictive multimodal fusion within the framework, healthcare professionals can proactively identify individuals at higher risk of developing specific diseases or conditions, facilitating preventive interventions and personalized healthcare strategies~\cite{acosta2022multimodal, nabbout2020impact}. For instance, early identification of individuals at risk of cardiovascular disease enables targeted lifestyle modifications, medication interventions, and regular monitoring to prevent or manage the condition~\cite{tao2021remote}. Through the integration of multimodal data and the application of predictive analytics, the Predictive component enhances healthcare decision-making and supports proactive interventions, ultimately improving patient outcomes and healthcare delivery~\cite{shaik2022ai}.

\subsection{Preventive Healthcare}
In the DIKW framework, the ``Preventive'' component of multimodal fusion focuses on utilizing diverse data sources to develop personalized preventive strategies for patients. By integrating and analyzing data from various sources, such as mHealth devices and EHRs, healthcare providers can gain insights into patients' lifestyle factors and health issues, enabling them to develop targeted interventions and preventive measures~\cite{nabbout2020impact, boehm2022harnessing}. By combining data from sources such as mHealth devices and EHRs, healthcare providers can gain a comprehensive understanding of patients' health status and develop personalized preventive strategies~\cite{liefaard2021way, muhammad2021comprehensive}.

Multimodal data fusion enables the integration and analysis of data from diverse sources, such as wearable fitness trackers or smart watches for monitoring physical activity, heart rate, and sleep patterns, and EHRs containing medical history, diagnoses, and laboratory results~\cite{shaik2022fedstack}. By combining these data sources, healthcare providers obtain a holistic view of patients' health and lifestyle factors~\cite{naqvi2020insights}. Through multimodal fusion, healthcare providers can identify lifestyle factors that may contribute to patients' health issues. Data from mHealth devices and EHRs may indicate emergent patterns such as sedentary behavior, inadequate sleep, or poor nutrition that may impact the development or progression of certain health conditions~\cite{zhang2020advances, bi2021novel, vale2021long}. 

Based on the analysis of multimodal data, healthcare providers can develop personalized preventive strategies tailored to each patient's unique needs. For instance, if data fusion analysis reveals that a patient's sedentary behavior contributes to their health issues, healthcare providers may prescribe regular physical activity, provide educational materials on exercise, or suggest behavioral change techniques to promote a more active lifestyle~\cite{horgan2022accelerating}. Furthermore, multimodal fusion facilitates ongoing monitoring and feedback, empowering patients to maintain their preventive strategies and make informed decisions about their health. By leveraging technology and data integration, patients can receive real-time feedback on their health behaviors, track their progress, and receive personalized recommendations to support their preventive efforts~\cite{cai2020feature, mateo2022delivering}. By incorporating the Preventive component within the DIKW and generic frameworks, multimodal fusion plays a vital role in developing personalized preventive strategies for patients, aligning with the broader goals of precision medicine and personalized healthcare~\cite{aceto2020industry, ruiz2022artificial}.

\subsection{Personalized Healthcare}
In the DIKW framework, the ``Personalized'' component of multimodal fusion focuses on utilizing diverse data sources to develop personalized treatment plans for patients. By integrating and analyzing data from different modalities, such as imaging and genomics, healthcare providers can gain a deeper understanding of the patient's molecular profile and tailor treatment strategies based on their individual characteristics~\cite{nabbout2020impact, boehm2022harnessing}. 

Within the generic framework, the Personalized component of multimodal fusion aims to identify specific genetic mutations or variations that underpin the patient's disease. By combining imaging data with genomics data, healthcare providers can obtain a comprehensive view of the patient's health condition and uncover genetic markers associated with conditions such as tumor growth~\cite{liefaard2021way, muhammad2021comprehensive}. Multimodal data fusion allows for the integration and analysis of data from diverse sources, such as magnetic resonance imaging (MRI), computed tomography (CT), positron emission tomography (PET), and genomics data. These modalities provide detailed anatomical, functional, and genetic information about the patient's body and disease state~\cite{naqvi2020insights}. 

By fusing imaging and genomics data, healthcare providers can identify specific genetic mutations or variations that inform the underlying molecular mechanisms of the disease. This knowledge guides the development of personalized treatment plans tailored to the patient's individual molecular profile~\cite{horgan2022accelerating}. The analysis of multimodal fusion enables healthcare providers to make informed treatment decisions, such as recommending targeted therapies for specific genetic mutations. This personalized approach ensures that patients receive the most effective treatments based on their unique genetic characteristics~\cite{cai2020feature}. 

Moreover, multimodal fusion facilitates ongoing monitoring of treatment response and enables treatment adjustments over time. By integrating data from imaging, genomics, and other modalities, healthcare providers can assess treatment effectiveness and make informed decisions regarding treatment modifications to optimize patient outcomes~\cite{mateo2022delivering}. By incorporating the Personalized component within the DIKW and generic frameworks, multimodal fusion can play a significant role in tailoring treatment plans based on individual patient characteristics and molecular profiles. This personalized approach to patient care and treatment outcomes aligna with the broader goals of precision medicine and personalized healthcare~\cite{aceto2020industry}.

\subsection{Participatory Healthcare}
In the DIKW framework, the ``Participatory'' component of multimodal fusion in smart healthcare focuses on empowering patients to actively participate in their own healthcare journey, fostering collaboration and shared decision-making with healthcare providers~\cite{boehm2022multimodal, carayon2020seips}. Within the DIKW framework, the Participatory component involves leveraging multimodal fusion to provide patients with real-time feedback and insights into their health status~\cite{dhayne2019search}. By integrating data from mHealth devices and patient-reported information, patients can actively monitor and track their health, enabling them to make informed decisions about their well-being~\cite{el2021potential}. 

In the generic framework, the Participatory component of multimodal fusion emphasizes the active engagement of patients in their healthcare by combining data from mHealth devices and patient-reported data~\cite{walker2023quality}. Through real-time access to their health information, patients can receive personalized feedback and recommendations, and participate in discussions about their treatment plans ~\cite{boehm2022multimodal}. This collaborative approach enables patients to actively contribute to decision-making based on their preferences, values, and personal health goals. Multimodal data fusion also enables patients to participate in larger-scale initiatives, such as contributing their data to aggregated and anonymized datasets~\cite{carayon2020seips}. By participating in research studies, clinical trials, or public health monitoring programs, individuals can contribute to advancements in medical research, personalized interventions, and population health initiatives. By embracing the Participatory component of multimodal fusion, patients become active partners in their own healthcare, and can be further empowered to make proactive choices and actively contribute to their own well-being. This collaborative approach enhances patient-centered care and fosters a stronger partnership between patients and healthcare providers.

\section{Conclusion}\label{conclusion}

Multimodal medical data fusion, integrating various modalities like EHRs, medical imaging, wearable devices, genomic data, sensor data, environmental data, and behavioral data, has the potential to revolutionize smart healthcare. By leveraging approaches such as feature selection, rule-based systems, ML, deep learning, and NL, practitioners can extract valuable insights from a wealth of diverse sources, which will advance gains in knowledge and wisdom in healthcare. 

However, the challenges related to data quality, interoperability, privacy, security, data processing, clinical integration, and ethical considerations must be addressed. Future research should focus on Predictive, Preventive, Personalized, and Participatory approaches, the implementation or combination of which which can enable better anticipation of health events, identify risk factors, deliver tailored interventions, or further empower patients in their healthcare journeys. Embracing these opportunities will transform healthcare by improving patient well-being, treatment outcomes, and the overall function of the healthcare industry.

\section*{Acknowledgement}
The authors gratefully acknowledge financial support  ANID Fondecyt 1231122, PIA/PUENTE AFB220003,  Chile.

\footnotesize
\bibliographystyle{ieeetr}

\bibliography{ieee}

\end{document}